%%%%%%%%%%%%%%%%%%%%%%%%%%%%%%%%%%%%%%%%%%%%%%%%%%%%%%%
% filename="Esing.tex"

%%%% version 20th May, 2001

\input harvmac

%%%%%%%%%%%%%%%%%%%%%%%%%%%%%%%%%%%%%%%%%
% Basic definitions
%%%%%%%%%%%%%%%%%%%%%%%%%%%%%%%%%%%%%%%%%

%

\def\coeff#1#2{\relax{\textstyle {#1 \over #2}}\displaystyle}
\def\xcoef(#1,#2){\coeff{#1}{#2}\,}
\def\half{{1 \over 2}}
\def\LG{Lan\-dau-Ginz\-burg\ }

 \def\cI{{\cal I}}

\def\cN{{\cal N}} 
 
\def\cR{{\cal R}} \def\cS{{\cal S}}
 
 \def\cW{{\cal W}}
\def\cX{{\cal X}} 
\def\ie{{\it i.e.}}
\def\bfone{\relax{\rm 1\kern-.35em 1}}
\def\inbar{\vrule height1.55ex width.4pt depth0pt}
\def\IC{\relax\,\hbox{$\inbar\kern-.3em{\rm C}$}}
\def\ID{\relax{\rm I\kern-.18em D}}
\def\IF{\relax{\rm I\kern-.18em F}}
\def\IH{\relax{\rm I\kern-.18em H}}
\def\II{\relax{\rm I\kern-.17em I}}
\def\IN{\relax{\rm I\kern-.18em N}}
\def\IP{\relax{\rm I\kern-.18em P}}
\def\IQ{\relax\,\hbox{$\inbar\kern-.3em{\rm Q}$}}
\def\IR{\relax{\rm I\kern-.18em R}}
\def\us#1{\underline{#1}}
\font\cmss=cmss10 \font\cmsss=cmss10 at 7pt
\def\ZZ{\relax\ifmmode\mathchoice
{\hbox{\cmss Z\kern-.4em Z}} {\hbox{\cmss Z\kern-.4em Z}}
{\lower.9pt\hbox{\cmsss Z\kern-.4em Z}}
{\lower1.2pt\hbox{\cmsss Z\kern-.4em Z}}\else{\cmss Z\kern-.4em Z}\fi}

\def\nihil#1{{\it #1}}
\def\eprt#1{{\tt #1}}
\def\nup#1({Nucl.\ Phys.\ $\us {B#1}$\ (}
\def\plt#1({Phys.\ Lett.\ $\us  {#1B}$\ (}
\def\cmp#1({Comm.\ Math.\ Phys.\ $\us  {#1}$\ (}
\def\prp#1({Phys.\ Rep.\ $\us  {#1}$\ (}
\def\prl#1({Phys.\ Rev.\ Lett.\ $\us  {#1}$\ (}
\def\prv#1({Phys.\ Rev.\ $\us  {#1}$\ (}
\def\mpl#1({Mod.\ Phys.\ Let.\ $\us  {A#1}$\ (}
\def\ijmp#1({Int.\ J.\ Mod.\ Phys.\ $\us{A#1}$\ (}
\def\jag#1({Jour.\ Alg.\ Geom.\ $\us {#1}$\ (}

%

%%%%%%%%%%%%%%%%%%%%%%%%%%%%%%%%%%%%%%%%%

%%%%%%%%%%%%%%%%%%%%%%%%%%%%%%%%%%%%%%%%%
% References
%%%%%%%%%%%%%%%%%%%%%%%%%%%%%%%%%%%%%%%%%
%
\lref\Catast{E.~Martinec, \nihil{Algebraic Geometry and
Effective Lagrangians,} \plt{217} (1989) 431; \hfil\break
C.~Vafa and N.P.~Warner, \nihil{Catastrophes and the Classification
of  Conformal Theories,} \plt{218} (1989) 51.}
\lref\SWStrings{S.\ Kachru, A.\ Klemm, W.\ Lerche, P.\ Mayr and
C.\ Vafa, \nihil{Non-perturbative Results on the Point Particle
Limit of N=2 Heterotic String Compactifications,} \nup{459}
(1996) 537, hep-th/9508155; \hfil\break
A.~Klemm, W.~Lerche, P.~Mayr, C.~Vafa and N.P.~ Warner
\nihil{Self-Dual Strings and N=2 Supersymmetric Field Theory,}
\nup{477} (1996) 746,  hep-th/9604034}
\lref\TopMat{E.~Witten,
\nihil{Topological Quantum Field Theory,}
\cmp{117} (1988)  353; \hfil \break
T.~Eguchi and S.-K.~Yang,
\nihil{N=2 Superconformal Models As Topological Field Theories,}
\mpl{5} (1990) 1693.}
\lref\TopGr{R.~Dijkgraaf, H.~Verlinde and E.~Verlinde,
\nihil{Topological Strings in $D < 1$,} \nup{352} (1991) 59; \hfil\break
K.~Li, \nihil{Recursion Relations in Topological Gravity with Minimal
Matter,} \nup{354} (1991)  725 ; \hfil \break
T.~Eguchi, Y.~Yamada and S.-K.~Yang,
\nihil{Topological field theories and the period integrals,}
\mpl{8} (1993) 1627,  hep-th/9304121;  \hfil\break
T.~Eguchi, H.~Kanno, Y.~Yamada and S.-K.~Yang,
\nihil{Topological Strings, Flat Coordinates and Gravitational
Descendants,} \plt{305}  (1993) 235, hep-th/9302048.}
\lref\TESKYa{ T.~Eguchi and S.-K.~Yang, \nihil{A New Description of
the $E_6$ Singularity,} \plt{394} (1997) 315;
\eprt{hep-th/9612086}.}
\lref\NSEWa{N.\ Seiberg and E.\ Witten, \nihil{Electric-Magnetic Duality,
Monopole Condensation, and Confinement in N=2 Supersymmetric
Yang-Mills Theory}, \nup426(1994) 19, hep-th/9407087.}
\lref\NSEWb{N.\ Seiberg and E.\ Witten,\nihil{Monopoles, Duality and
Chiral Symmetry Breaking in N=2 Supersymmetric QCD}, \nup426(1994) 19,
\nup{431} (1994) 484, hep-th/9408099.}
\lref\GVW{S.~Gukov, C.~Vafa and E.~Witten,
\nihil{CFT's From Calabi-Yau Four-folds,} \nup{584} (2000) 69;
\eprt{hep-th/9906070}.}
\lref\Beckers{K.~Becker and M.~Becker,
\nihil{M-Theory on Eight-Manifolds,} \nup{477} (1996) 155;
\eprt{hep-th/9605053}.}
\lref\LVW{W.~Lerche, C.~Vafa and N.~P.~Warner,
\nihil{Chiral Rings in $N=2$ Superconformal Theories,}
\nup{324} (1989) 427.}
\lref\LWpoly{W.~Lerche and N.~P.~Warner,
\nihil{Polytopes and Solitons in Integrable, $N=2$ Supersymmetric
Landau-Ginzburg Theories,} \nup{358} (1991)  571.}
\lref\ASNW{A.N.~Schellekens and N.~P.~Warner,
\nihil{Conformal Subalgebras of Kac-Moody Algebras,}
\prv{D34} (1986) 3092.}
\lref\DNNW{D.~Nemeschansky and N.~P.~Warner,
\nihil{Refining the Elliptic Genus,}
\plt{329} (1994) 53; \eprt{hep-th/9403047}.}
\lref\WLAS{W.~Lerche and A.~Sevrin,
\nihil{On the Landau-Ginzburg Realization of Topological Gravities,}
\nup{428} (1994) 259; \eprt{hep-th/9403183}.}
\lref\WLNWEsix{ W.~Lerche and N.~P.~Warner, \nihil{Exceptional
SW geometry from ALE fibrations,} \plt{423} (1998) 79;
\eprt{hep-th/9608183}.}
\lref\FracFerm{R.~Jackiw and C.~Rebbi, \nihil{Solitons with Fermion
Number 1/2,} \prv{D13} (1976) 3398; \hfil \break
J.~Goldstone and F.~Wilczek, \nihil{Fractional Quantum
Numbers on Solitons,} \prl{47} (1981) 986 .}
\lref\PFKI{ P.~Fendley and K.~Intriligator, \nihil{Scattering and
Thermodynamics of Fractionally-Charged Supersymmetric Solitons,}
\nup{372} (1992) 533;  \eprt{hep-th/9111014}; \nihil{Scattering and
Thermodynamics of Integrable $N=2$ Theories,}
\nup{380} (1992) 265; \eprt{hep-th/9202011}.}
\lref\ItYa{K.~Ito and S.-K.~Yang,
\nihil{Flat Coordinates, Topological Landau-Ginzburg Models and the
Seiberg-Witten Period Integrals,}
\plt{415} (1997) 45; \eprt{hep-th/9708017};
\nihil{A-D-E Singularity and Prepotentials in N=2 Supersymmetric Yang-Mills
Theory,}
\ijmp{13} (1998) 5373; \eprt{hep-th/9712018}.}
\lref\EKMY{T.~Eguchi, T.~Kawai, S.~Mizoguchi and S.-K.~Yang,
\nihil{Character Formulas for Coset N=2 Superconformal Theories,}
Rev.\ Math.\ Phys.\ $\us{4}$ (1992) 329.}
\lref\KazSuz{Y.~Kazama and H.~Suzuki, \nihil{Characterization of N=2
Superconformal Models Generated by Coset Space Method,} \plt{216} (1989)
112; \nihil{New N=2 Superconformal Field Theories and Superstring
Compactification} \nup{321} (1989) 232.}
\lref\EMNW{E.~Martinec and N.P.~Warner, \nihil{Integrable systems
and supersymmetric gauge theory,} \nup{459} (1996) 97, hep-th/9509161}
\lref\IntLG{P.~Fendley, S.~D.~Mathur, C.~Vafa and N.~P.~Warner,
\nihil{Integrable Deformations and Scattering Matrices for the N=2
Supersymmetric Discrete Series,} \plt{243} (1990) 257; \hfil \break
 A.~LeClair, D.~Nemeschansky and N.~P.~Warner,
\nihil{S-matrices for Perturbed N=2 Superconformal Field
Theory from Quantum Groups,}  \nup{390} (1993)  653,
\eprt{hep-th/9206041}.}
\lref\Entori{J.A.~Minahan and D.~Nemeschansky, \nihil{Superconformal
Fixed Points with $E_n$ Global Symmetry,} \nup{489} (1997) 24,
\eprt{hep-th/9610076}; \hfil \break
J.~A.~Minahan, D.~Nemeschansky and N.~P.~Warner,
\nihil{Investigating the BPS spectrum of Non-Critical $E_n$ Strings,}
\nup{508} (1997) 64,  \eprt{hep-th/9705237}.}
\lref\DVV{R.~Dijkgraaf, E.~Verlinde and H.~Verlinde,
\nihil{Topological Strings in $D< 1$,} \nup{352} (1991) 59.}
%
%%%%%%%%%%%%%%%%%%%%%%%%%%%%%%%%%%%%%%%%%

%%%%%%%%%%%%%%%%%%%%%%%%%%%%%%%%%%%%%%%%%
% Title
%%%%%%%%%%%%%%%%%%%%%%%%%%%%%%%%%%%%%%%%%
\Title{ \vbox{ \hbox{CITUSC/01-017} \hbox{USC-01/01} \hbox{\tt
hep-th/0105194} }} {\vbox{\vskip -1.0cm
\centerline{\hbox
{ADE Singularities and Coset Models}}}}
\vskip -.3cm
\centerline{Tohru Eguchi$^*$,  Nicholas P.\ Warner$^\dagger$ and
Sung-Kil Yang$^{\dagger\dagger}$}
\bigskip
\bigskip
\centerline{$^*${\it Department of Physics,
University of Tokyo, Tokyo 113-0033, Japan}}
\medskip
\centerline{$^\dagger${\it Department of Physics and Astronomy
and  CIT-USC Center for Theoretical Physics}}
\centerline{{\it University of Southern California,
 Los Angeles, CA 90089-0484, USA}}
\medskip
\centerline{$^{\dagger\dagger}${\it Institute of Physics,
University of Tsukuba, Ibaraki 305-8571, Japan}}
%%%%%%%%%%%%%%%%%%%%%%%%%%%%%%%%%%%%%%%%%

%%%%%%%%%%%%%%%%%%%%%%%%%%%%%%%%%%%%%%%%%
% Abstract
%%%%%%%%%%%%%%%%%%%%%%%%%%%%%%%%%%%%%%%%%
\bigskip
\bigskip

We consider the compactification of the IIA string to
$(1+1)$ dimensions on non-compact $4$-folds that are
ALE fibrations.  Supersymmetry requires that the compactification
include $4$-form fluxes, and a particular class of these
models has been argued by Gukov, Vafa and Witten to give
rise to a set of perturbed superconformal coset models
that also have a Landau-Ginzburg description.   We examine
all these ADE models in detail, including the exceptional
cosets.  We identify which perturbations are induced by
the deformation of the singularity, and compute the
Landau-Ginzburg potentials exactly.  We also show how the
 the Landau-Ginzburg fields and their superpotentials
arise from the geometric data of the singularity, and we find that
this is most naturally described in terms of non-compact, holomorphic
$4$-cycles.

\vskip .3in
%\draft
\Date{\sl {May, 2001}}
%%%%%%%%%%%%%%%%%%%%%%%%%%%%%%%%%%%%%%%%%

%%%%%%%%%%%%%%%%%%%%%%%%%%%%%%%%%%%%%%%%%
% Body
%%%%%%%%%%%%%%%%%%%%%%%%%%%%%%%%%%%%%%%%%

%%%%%%%%%%%%%%%%%%%%%%%%%%%%%%%%%%%%%%%%%
\newsec{Introduction}
%%%%%%%%%%%%%%%%%%%%%%%%%%%%%%%%%%%%%%%%%
The link between the classification of singularity types
and quantum effective actions of field theories now has a
fairly long and interesting history.  The key to making
this work is for the field theory to have enough supersymmetry
so that a non-renormalization ``theorem'' protects the
sector of the field theory that is determined by the
classical singularity type.  For theories with a mass
gap there are generically BPS states whose spectrum can be
computed semi-classically, and which can be used to identify
the field theory.  This was first used with
considerable success in $\cN=2$ superconformal field
theories, in which there is a Landau-Ginzburg
superpotential \Catast.  Most particularly the ADE classification
of complex singularities with no moduli was shown to correspond to precisely
the modular invariant $\cN=2$ superconformal field theories
 with central charge $c < 3$.
If one topologically twists these $\cN=2$ theories
the result is topological matter \TopMat, and the coupling to
topological gravity could be naturally incorporated
into the singularity theory \refs{\TopGr,\TESKYa}.

The advent of string dualities gave rise to many new constructions
of string and field theories, and in particular, low energy effective
actions. An early offshoot of this general program was to realize that
the complete Seiberg-Witten quantum effective action of $\cN=2$ 
super-Yang-Mills theories in four dimensions \refs{\NSEWa,\NSEWb} could be
obtained from the period integrals of $K3$-fibrations that were developing
an $ADE$ singularity \SWStrings. In this construction, the singularity
type corresponds to the gauge group of the  Yang-Mills theory.

This subject has now, in a sense, come full circle.  In \GVW\ it was
shown how Landau-Ginzburg models in $(1+1)$-dimensions could be constructed
using compactifications of type IIA strings on Calabi-Yau $4$-folds.
If the $4$-fold in this construction is a $K3$-fibration, and if
the fiber develops an $ADE$ singularity, then the $(1+1)$-dimensional
field theory has been argued to be a massive perturbation of a
coset model based upon the corresponding $ADE$ group.
The mass scale is set by the deformation of the singularity,
and a conformal field theory emerges when the fiber is
singular.  

Extracting field theories from singular limits of compactifications
is often a subtle process, and there are sometimes 
multiple limits being taken at the same time.  For example,  
to extract the Seiberg-Witten effective actions one needs
to decouple gravity and yet maintain a cut-off whose dimensions 
are inherited from the string tension.  This is done 
by scaling the size of the base of the fibration while 
scaling the $K3$ so as to isolate the singular fiber.  This
results in a non-trivial role for the fibration even
in the field theory limit.   A conformal field theory can 
occur in this context  if the moduli of the singularity are 
fine-tuned to  an Argyres-Douglas point, but the generic theory 
has a scale.  Things are considerably
simpler for the \LG models that arise
from singular $K3$ fibers in $4$-folds:  there is no
cut-off, and the field theory is conformal when the 
$K3$ fiber is singular.  All the essential physics comes from the 
isolated $ADE$ singularity in the fiber, and the base of
the fibration only plays a trivial role.

While the singularity type determines the numerator of the
coset conformal field theory, the denominator (and probably the level)
of the model are determined by a background $4$-form flux.  Such fluxes
are parameterized by weights of the Lie algebra, $G$, associated
with the $ADE$ singularity.  The Weyl group of the singularity acts on
the  flux, permuting it around a Weyl orbit.  The denominator of the
coset model is defined  by the subgroup of the Weyl group that leaves a flux
invariant,  and the non-trivial Weyl images of each flux represent different
ground  states of the same  model.   Deformations of the singularity
introduce masses, but  in general the deformed theory
does not have a mass gap: there are still massless excitations.
However, for a certain class of minimal  fluxes\foot{A  minimal 
flux is one that corresponds to a miniscule weight of $G$.},  the theory
does have a mass gap, and  the associated coset model  was argued in
\GVW\ to be (a perturbation of) the  $\cN=2$, Kazama-Suzuki coset model
based  upon the level one, $A$, $D$ or $E$
hermitian symmetric space.  This was checked in detail
in \GVW\ for the $A$-type models, and here we will verify that
the construction works for the $D$-type and the $E$-type  singularities.
It is not immediately clear what the corresponding models are
for larger (non-minimal) fluxes, but it is tempting to try to
identify the magnitude of the flux with the level of the coset
model.

The identification of these models, even at level one, is a non-trivial
problem.  If one uses a non-compact Calabi-Yau
$4$-fold with an $ADE$ singularity as outlined above, it turns out
that all the dynamics are frozen because the kinetic terms of the
the model are not normalizable.  On the other hand, in \GVW\
a \LG superpotential, $W$, was conjectured for compact Calabi-Yau
$4$-folds, and for non-compact Calabi-Yau manifolds this yields
expressions for the topological charges of solitons  in terms of period
integrals on the $4$-fold.  The problem is that for non-compact
Calabi-Yau manifolds, there does not appear to be an obvious geometric
characterization of the \LG fields themselves: one does not,
{\it a priori,} know how many fields there are, let alone their $U(1)$
charges.    Thus the singularity only yields topological data
about the model, that is, one knows only the ground states and the
topological
charges of the solitons.  On the other hand, knowing the coset conformal
field
theory determines the chiral primary fields and the \LG potential
(if there is one).  Such a theory also generically has many more
deformations of the superpotential than there are deformations of
the singularity of the $4$-fold.  Thus there is a special, ``canonically''
deformed \LG superpotential whose deformations are precisely
generated from the deformations of the $4$-fold.

One of our purposes here is to characterize and compute all the
canonically deformed  superpotentials associated with the ADE
singularities with minimal  fluxes.
We first do this rather abstractly by developing
an algorithm for computing such superpotentials.  In so doing, we
find that there are obvious choices in the procedure, and that these choices
correspond to determining the coset denominator, and hence the correct set
of \LG variables.  We then look at the period integrals on the
$4$-fold, and find how to determine the superpotentials from such
calculations.  A direct computation appears to lead to only one
superpotential
for each $A$, $D$, or $E$ singularity.  However, we also show that there
are, once again, ambiguities in the calculation of the periods, and
that different choices lead to the full set of
canonically deformed superpotentials.  We will show that the
resolution of the ambiguity amounts to selecting holomorphic representatives
of the non-compact homology of the singularity.  These representatives
are characterized by  weights of $G$, and each choice is equivalent to
selecting a Weyl orbit of fluxes.  In doing the calculations
of the period integrals, we also find the vestigial remnants of
the \LG variables, and this leads to some natural conjectures as to
the role of the \LG variables in the geometric picture on $4$-folds.

The next two sections of this paper contain a detailed review of
the $\cN=2$ superconformal coset models of  Kazama and Suzuki, and
the construction of associated \LG superpotentials.  In particular,
in section 2 we will review that standard construction of
the chiral rings and \LG potentials of the simply laced, level
one, Hermitian symmetric space (SLOHSS) models.  We then generalize this
to obtain the  ``canonically'' deformed superpotentials.
In section 3 we will construct some specific examples of
these deformed superpotentials, and evolve an algorithm for
generating all the superpotentials associated with the numerator
 Lie algebra, $G$, from any one such superpotential.   In section 4
we first review the part of \GVW\ that is relevant to the SLOHSS  models,
and then use the formula of \GVW\ to calculate topological charges
of solitons in terms of period integrals of the singularity.   We then
use this calculation, in combination with the algorithm developed in
section 3, to reconstruct the \LG superpotentials from the geometry
of the singularity.     Our methods of computing the period integrals
can be more naturally cast in terms excising holomorphic surfaces, and
computing intersection numbers.  We also discuss this in section 4,
and show how the selection of the holomorphic surfaces is equivalent
to choosing a flux at infinity.  We also find that the \LG variables
appear very naturally in the parameterization of these surfaces.
Finally, in section 5 we make some remarks about fermion numbers
of solitons, and about generalizations of the results presented here.

\newsec{The Landau-Ginzburg Potentials of Deformed SLOHSS Models}

\subsec{The conformal Landau-Ginzburg potential}

The $\cN=2$, Kazama-Suzuki conformal coset models \KazSuz\ have been
extensively studied.  They are based upon the coset construction
using:
\eqn\KScoset{\cS ~=~ {G \times SO(dim(G/H)) \over H}\,,}
where the $SO(dim(G/H))$ is a level one current algebra, and represents
the bosonized fermions in a supersymmetrization of the coset model based
upon $G/H$.  In \KScoset, $H$ is embedded diagonally into
$G$ and  $SO(dim(G/H))$, the embedding into $G$ is an index
one embedding, and the embedding into $SO(dim(G/H))$ is
a conformal embedding \ASNW\ of level $g-h$ where $g$
and $h$ are the dual Coxeter numbers of $G$ and $H$ respectively.
If $G$ has a level $k$, then $H$ has a level $k + g - h$.
If $G/H$ is K\"ahler, then $H=H_0 \times U(1)$, with
$U(1)$ inducing the complex structure, and the
corresponding coset model has an $\cN=2$ superconformal
algebra.

The simply laced, level one, hermitian symmetric space (SLOHSS)
models have been studied even more extensively.  In these models, $G$
is represented by a level one current algebra, and $G/H$ is a hermitian
symmetric space.  These cosets are:
\eqn\HSSlist{\eqalign{ & {SU(n+m)\over SU(n)\times SU(m)\times U(1)}
\,,   \qquad {SO(2n)\over SO(2n-2)\times U(1) }  \,,\qquad
{SO(2n)\over SU(n) \times U(1)} \,, \cr &{E_6\over SO(10)\times U(1)}
\,,
\qquad {E_7\over E_6\times U(1)}\,.}}
It was shown in \LVW\ that the chiral ring of these coset models is
isomorphic
to the de Rham cohomology ring $H^*(\cS, \IR)$.  It has also been argued
that these models have a \LG formulation \LVW.  Most of the Landau-Ginzburg
potentials have been determined.

The most direct way to compute the chiral ring and potential
is to use the isomorphism with the cohomology ring of
$\cS$. The latter can be generated by the Chern classes of
$H$-bundles on $\cS$, and these can be generated from the irreducible
$H$-representations.  The vanishing relations
of the ring are characterized by the trivial $H$-bundles,
and the corresponding vanishing Chern classes
are given by those  combinations of $H$-representations
that are actually $G$-representations, and hence generate
trivial bundles.  In more mechanistic terms, the chiral ring
is generated by the Casimir invariants of $H$, and to
find the vanishing relations one simply has to take all the
Casimirs of $G$ and decompose them into the Casimirs of
$H$.

In practice it is simplest to reduce this calculation
to the Cartan subalgebra, $\cX$, of $G$.  (Since $G$ and $H$ have
the same rank, this is also a Cartan subalgebra (CSA) of
$H$.)  We then parameterize $\cX$ by variables, $\xi_j$,
$j =1,\dots,r$, where $r$ is the rank of $G$.
The Casimirs of $G$ and $H$ are then equivalent
to $W(G)$ and $W(H)$ invariant polynomials, denoted $V_j$ and
$x_i$ respectively,
of the $\xi_j$,  where $W(G)$ and $W(H)$ are the Weyl groups
of $G$ and $H$ respectively.   The inequivalent
$W(H)$ invariant polynomials, modulo $W(G)$ invariant polynomials,
are given by elements of $W(G)/W(H)$, and thus the chiral ring has
the  structure of this Weyl coset.

One should also observe that elements of the coset $W(G)/W(H)$, and
hence the ground states, are in one-to-one correspondence
with the weights of a ``miniscule'' representation of $G$.
A miniscule representation,  $\cR$, is defined as one in which all the
weights lie in a single Weyl orbit.  The subgroup $H$ is then
defined to be the one whose semi-simple part (not the $U(1)$)
fixes the highest weight of $\cR$.  Since all other weights lie
in the Weyl orbit of the highest weight, it follows that
the representation is indeed in one-to-one correspondence with
$W(G)/W(H)$. In this manner one can generate the list \HSSlist.
We thus see that ground states of the SLOHSS models are naturally
labeled by the weights of a miniscule representation of $G$.  This
will be important throughout this paper.

The ``reduction of Casimirs'' procedure was used in
\refs{\LVW,\LWpoly} to construct
the chiral rings.  The remarkable, and rather unexpected fact
is that the vanishing relations appear to be integrable
to produce a single superpotential for each SLOHSS model
\foot{This has not yet been proven for the general coset
${SO(2n)\over SU(n) \times U(1)}$, but in fact it will
be implicitly established in this paper.}.  In particular,
the Landau-Ginzburg potentials for $E_6$ and $E_7$ were
computed in \LWpoly.

The basic procedure is therefore as follows:  Let  $m_i$ and
$\widehat m_i$, $i=1,\dots, r$, be the exponents of $G$ and $H$
respectively. (We define the exponent of a $U(1)$
to be zero.)  The degrees of the Casimirs are
thus $m_i+1$ and $\widehat m_i +1$. (We define the Casimir
of $U(1)$ by taking a trace: it thus has degree $1$.)
The generators of the chiral ring are thus
variables $x_i$ , $i=1,\dots, r$, of degrees
$\widehat m_i +1$.  The vanishing relations are polynomials
$V_i(x_j)$,  $i=1,\dots, r$, of degrees
$m_i +1$, obtained by the decomposition of Casimir
invariants.  To get the Landau-Ginzburg potential
one wants to integrate the vanishing relations
to obtain a potential $W(x_i)$ such that the
set of equations  $V_i = 0$ are equivalent to
the set of equations ${\del W \over \del x_i} = 0$.
The only complexity in this task is that
$\del W \over \del x_j$ will generically be some
constant multiple of $V_j$ plus $V_k$'s
of lower degree multiplied by polynomials
in the $x_i$'s.  One has thus to find the proper
combinations of $V_k$'s before integration is possible,
and as we indicated earlier, it seems remarkable that
such integration is possible.

\subsec{The canonically deformed Landau-Ginzburg potential}

The conformal \LG potential is, of course, quasi-homogeneous,
and therefore multi-critical.  The index of the
singularity, $\mu = \Big|{W(G) \over W(H)} \Big|$, is
the degeneracy of the Ramond ground states.  We now wish
to deform this potential in such a manner as to make a
mass gap, and yield the corresponding theories associated
with the $ADE$ singularities in \GVW.  The key to seeing
how this must happen is to observe that the versal deformation
of the singularity preserves the $W(G)$ symmetry of the
singularity, and therefore so must the corresponding
deformation of the coset model.  In particular, the ground
states of the deformed model must be a $W(G)$-invariant
family.  This leads to a unique versal deformation procedure
for the coset model.
Instead of setting the $G$-Casimirs, $V_i(x_j)$, to zero
one can set them to constant values: $V_i = v_i$ for some $v_i$.
This is manifestly $W(G)$ invariant, and represents a set
of equations that define the ground states of the canonically
deformed model.

To understand the interrelationship between all the \LG potentials
for cosets $G/H$ with the same numerator $G$,
but with different denominators $H$, it
is important to understand how these ground states, and hence
the canonical deformation,
is realized in terms of the CSA variables, $\xi_j$.
In terms of the Cartan subalgebra, $\cX$,
setting $V_i = v_i$ uniquely defines a general point, $\xi_j
= \xi_j^{(0)}$, in $\cX$ up to the action of $W(G)$.
That is, the $v_j$ define a general point, $\xi_j^{(0)}$,
in the fundamental Weyl chamber of $G$.  Similarly, the values
of  the Casimirs, $x_j$, of $H$ uniquely specify a point
in the fundamental Weyl chamber of $H$.  The vacua of the
deformed coset model are thus characterized by all the $W(G)$
images of $\xi_j^{(0)}$ that lie in the fundamental Weyl chamber
of $H$. Non-zero values of the $v_j$  generically yield a massive
theory with all vacua separated and $\xi_j^{(0)}$ an interior point of
the Weyl chamber.  The theory becomes multi-critical, with
massless solitons when the $v_j$ assume values at which
$\xi_j^{(0)}$ goes to a wall of the Weyl chamber of $G$.

Remarkably enough, we find that the deformed ``vanishing relations:''
$V_i = v_i$ are still integrable to a Landau-Ginzburg potential
$W(x_j; v_i)$.   This will be proven in the next section.
Based upon the foregoing observation we develop an algorithm
for computing the desired \LG potential for any SLOHSS model,
$G/H$, given one such superpotential for any SLOHSS model with
the same numerator, $G$.  We then compute a superpotential
for each choice of $G$, including the exceptional cosets, and we also
give several examples of the application of the algorithm.

Finally, we note that the deformed potential is quasi-homogeneous:
\eqn\quasihom{W\big(\lambda^{\widehat m_j +1} \, x_j\, ;\
\lambda^{ m_i +1} \, v_i\,) ~=~ \lambda^{N+1}\,
W(x_j\,;\ v_i) \,,}
where $N= m_r$ is the dual Coxeter number ({\it i.e.}
the degree of the highest Casimir) of $G$.

Once again, we stress that because we have preserved the Weyl
symmetry of $G$, it is these deformed potentials  that must
be related to those of the versal deformations of the $ADE$ singularities.

\newsec{Chiral rings and superpotentials}

\subsec{Grassmannians}

Here one has $G= SU(m+n)$ and $H= SU(m) \times SU(n)
\times U(1)$, but it is simpler to think of this coset as having
$G= U(m+n)$ and $H= U(m) \times U(n)$.   We will take $ m \le n$.
The CSA of $U(m+n)$ can be parameterized by $(\xi_1,\dots, \xi_{m+n})$,
and the Casimirs of $G$ are the permutation invariants:
$$
V_k ~\equiv~ \sum_{\ell =1}^{m+n} \ \xi_\ell^k   \,, \quad
k=1,\dots, m+n
$$
while those of $H$ are:
\eqn\Hperminvs{x_k ~\equiv~ \sum_{\ell =1}^{m} \ \xi_\ell^k \,,
\quad k=1,\dots, m\,;  \quad \qquad  \tilde x_k ~\equiv~
\sum_{\ell =m+1}^{m+n} \ \xi_\ell^k \,,  \quad k=1,\dots, n \,.  }
It is convenient to introduce an equivalent set of Casimirs:
$$
\widehat V_k ~\equiv~ (-1)^k\sum_{1\le j_1<j_2< \dots <j_k \le m+n}  \
\xi_{j_1}\, \xi_{j_1}\, \dots  \xi_{j_k}  \,, \quad
k=1,\dots, m+n  \,,
$$
along with a corresponding family of new variables:
\eqn\zkdefn{ z_k ~\equiv~ (-1)^k\sum_{1\le j_1<j_2< \dots
<j_k \le m}  \  \xi_{j_1}\, \xi_{j_1}\, \dots  \xi_{j_k}
\,, \quad k=1,\dots, m \,. }

Since $x_k + \tilde x_k = V_k = v_k$, we can use this linear
equation to eliminate all the $\tilde x_k$ in terms of $x_k$ and
$v_k$.  Note that for $k > m$ one must write $x_k$ as a polynomial in
the $x_j$ ($j\le m$), and that for $k > n$ one must write $\tilde x_k$
in terms of a polynomial in the $\tilde x_j$ ($j\le n$).
Thus the deformed Landau-Ginzburg  potential is a function, $W_{m,n}$,
that depends upon $x_j$, $j=1, \dots, m$ and $v_k$, $k=1, \dots,m+n$.

For $m=1$ one can set $x=x_1= \xi_1$, and one has $v_{n+1} = x^{n+1}
+ \tilde x_{n+1}$.  One then writes $\tilde x_{n+1}$
as a polynomial in $\tilde x_j$, $j\le n$ and then eliminates the
$\tilde x_j$ using $\tilde x_k = v_k -x^k$. The result is
a superpotential of the form:
\eqn\Wbasic{ W ~\equiv~ W_{1,n} ~=~ \coeff{1}{n+2}\, x^{n+2} ~+~
\sum_{k=2}^{n+2} \coeff{1}{n+2-k}\ \hat v_k \  x^{n+2-k} \,,}
where $\hat  v_k$ are the values of $\widehat V_k$ given
$V_k = v_k$.  The whole point is that one has
\eqn\facdWdx{{d\,W \over d\,x } ~=~ \prod_{\ell=1}^{n+1}\
(x - \xi_{\ell}^{(0)}) \,.}
The $n+1$ critical points of this superpotential are then
given by  the components of the weight, $\xi_{\ell}^{(0)}$,
corresponding to the values, $v_k$, of the Casimirs, $V_k$.

Our purpose now is to show how to generate the superpotentials
of all the Grassmannians from \Wbasic.
For general, $m$, the chiral ring is generated by the, $x_k$, of
\Hperminvs\ which are permutation invariants of the first $m$
generators $\xi_1,\dots,\xi_m$.  Then there are $\left({m+n
\atop m}\right)$ ground states given by all choices of $m$
of the $\xi_{\ell}^{(0)}$.  In particular, the lowest dimension
chiral primary, $ x_1 \equiv \xi_1 + \dots + \xi_m$ takes
values $\xi_{j_1}^{(0)} + \dots+ \xi_{j_m}^{(0)}$ for all choices
of $j_1,\dots,j_m$.

One can thus extract the multi-variable potential from \Wbasic\
by a rather simple algorithm.  The ground states of $W_{m,n}$
are characterized by subsets of $m$ of the solutions to
${d W_{1,n} \over d x} = 0$.  The idea is to introduce
an auxiliary equation that defines such a subset of $m$
roots, and then use cross elimination between \Wbasic\ or
\facdWdx\ and this auxiliary equation to reconstruct the
superpotential $W_{m,n}$.  While the general superpotentials
for the Grassmannians might be constructed more directly, the
beauty of the foregoing algorithm is that it generalizes to
other Lie algebras.

For the general Grassmannian model we need the $m$ variables,
$x_k$, or equivalently and more conveniently, the $z_k$ of \zkdefn.
These characterize a subset, $\xi_1, \dots, \xi_m$,
of the $m+n$ roots of ${d W_{1,n} \over d x} = 0$.  Moreover, from
\zkdefn\ it follows that  the individual roots, $\xi_1, \dots, \xi_m$,
are related to the $z_k$ as the $m$ roots of the auxiliary equation:
\eqn\xzreln{  x^m ~+~ \sum_{k=1}^{m} \ z_k \,x^{m-k}
~=~ \prod_{j=1}^{m}\ (x - \xi_j) ~=~ 0\,.}
Thus, given a critical point of $W_{m,n}$, one can use this
formula to extract the set of $m$ critical points,
$\{\xi_{j_1}^{(0)}, \dots, \xi_{j_m}^{(0)} \}$ of \Wbasic,
that characterize a single ground state of the Grassmannian model

One now reverses this procedure:  the multi-variable
critical points are precisely characterized by adjusting the
$z_k$ (or $x_k$) so that {\it all} $m$ roots
of \xzreln\ are critical points of \Wbasic.  The job is thus
to properly recast \Wbasic\ in terms of the $z_k$ using
the fact that we now wish to sum \Wbasic\ over  subsets of $m$
roots of ${d W_{1,n} \over d x} = 0$.

One can do this by using \xzreln\ to eliminate all powers, $x^k, k \ge m$,
in ${d W_{1,n} \over d x}$.  The result is a polynomial
$P(z_j;x)$ of overall degree $n+1$, but of degree $m-1$ in
$x$. If one makes the expansion: $P(z_j;x) =\sum_{\ell=0}^{m-1}
B_\ell(z_j)\, x^\ell$, then this will vanish for a (generic) set of
$m$ roots of ${d W_{1,n} \over d x} =0$
if and only if all of the $B_\ell(z_j)$ vanish.  These must
therefore be the deformed ``vanishing relations'' that
characterize the chiral ring of the multi-variable model.  It
is these that must be integrated to give the multi-variable
superpotential.

There is, however, a simpler way to get the \LG superpotential:
Perform the same elimination procedure, using \xzreln,
on the superpotential \Wbasic.  The result is once again
a function,  $W(z_j; x)$, that is a polynomial of degree
at most $m-1$ in $x$.   Now replace $x^j$ in this function
by ${1 \over m} x_j$, and rewrite everything in terms of either $z_j$
or $x_j$.  We claim that this results in the requisite
multi-variable potential, $W_{m,n+1-m}(x_j;v_j)$.  To see why this is
so, observe that this prescription for replacing $x^j$
is the same as summing ${1 \over m}  W(z_j; x)$ over the
$m$ roots of \xzreln.  Thus $W_{m,n+1-m}(x_j;v_j)$ represents an
average of the values of \Wbasic\ over a set of roots
of ${d W_{1,n} \over d x} =0$, and imposing ${\del W_{m,n+1-m}(x_j;v_j)
\over \del x_j} = 0$ implies that ${d W_{1,n} \over d x} =0$
on all $m$ of these roots.  Also recall that the values of the
superpotential encodes the topological charge of solitons, and
this averaging procedure reproduces the proper topological
charge in the multi-variable case.

To illustrate this procedure we consider the potentials for which
$m=2$.  Equation \xzreln\ implies:
\eqn\simpxz{ x~=~  \coeff{1}{2}\, z_1 ~\pm~ \coeff{1}{2}\,
\sqrt{z_1^2 ~-~ 4\, z_2} \,.}
One now substitutes this into \Wbasic\ and sums over both roots,
which amounts to dropping all square-roots from the result.
For $n=4$ and $5$ one obtains:
\eqn\sufivesix{\eqalign{  W_{2,3}(y,z) ~=~ &
\coeff{11}{12} \,z^6 - \coeff{1}{3}\,y^3 + \coeff{5}{4}\,y\,z^4
+ \coeff{1}{2}\,a_2 \,( y\,z^2 +  z^4) -  a_3\,( y\,z +
\coeff{7}{6} \,z^3) \cr &
- (a_4 - \coeff{1}{4}\,{a_2}^2)\,( y +   z^2)
+ ( a_5   - \coeff{1}{2} \,a_2\,a_3)\,z  +
\coeff{1}{12}\,{a_2}^3 - \coeff{1}{2}\,a_2\,a_4  \,, \cr
W_{2,4}(z_1,z_2)  ~=~ & \coeff{1}{7}\, z_1^7 - z_2^3\,z_1  +
2\,z_2^2\,z_1^3 -  z_2\,z_1^5 +  a_2\, \big( z_2^2\,z_1 -
z_2\,z_1^3 + \coeff{1}{5} \,z_1^5  \big) \cr & +
a_3\,\big(\coeff{1}{2}\,z_2^2 - z_2\,z_1^2 + \coeff{1}{4} \,z_1^4\big)
+  a_4 \, \big( \coeff{1}{3}\, z_1^3 -  z_2\,z_1\big) +  a_5 \,
\big( \coeff{1}{2}\, z_1^2 -z_2\big) +  a_6\, z_1  \,.}}
In the first of the superpotentials we have made the change of
variables: $z= z_1, z_2 = y + {3 \over 2} z^2 + {1 \over 2} a_2$.
Note, in particular that the value of the modulus
(the coefficient of $ y z^4$, with $y$ and $z$ suitably
normalized) in $W_{2,3}$ is consistent with the results of
\refs{\LWpoly,\DNNW,\WLAS}.

\subsec{The $SO(2n)$ superpotentials}

There are two infinite series of coset models with this numerator, and
we begin with the $SO(2n)/(SO(2n-2)\times U(1))$ coset model.
The denominator group, $SO(2n-2)\times U(1)$, leaves an $SO(2n)$
pair of vectors invariant, and so ground states of this model are
characterized by the weights of the vector representation of $SO(2n)$.
This model may also be thought of as the $D_{2n}$ minimal model, but we
will see that the canonical deformation leads to a subset of the full
set of deformations of the standard $D_{2n}$ potential.

The Casimirs of $SO(2n)$ are defined by
\eqn\SOcasi{\eqalign{
V_{2k} ~=~ & \sum_{1 \leq i_1 < i_2 <\cdots < i_k \leq n}
a_{i_1}^2\, a_{i_2}^2 \cdots a_{i_k}^2 \,,
\hskip10mm k=1,2,\dots,n-1 \,,   \cr
\widetilde V_n ~=~ & a_1\, a_2 \cdots a_n \,,  }}
where the $a_i$ are the skew eigenvalues of an $SO(2n)$ matrix
{\it in the vector representation}.
Under $SO(2n) \supset SO(2n-2)\times U(1)$ the Casimirs of $SO(2n)$ are
decomposed into the Casimirs $x_j$ of $SO(2n-2)\times U(1)$,
$$
\eqalign{
& V_2 ~=~ x_2+x_1^2 \,, \hskip10mm
V_{2j} ~=~ x_{2j} +x_{2j-2}\, x_1^2 \,; \hskip10mm j=2,3,\dots,n-2 \,, \cr
& V_{2n-2} ~=~ \tilde x_{n-1}^2+x_{2n-4}\, x_1^2 \,,
\hskip10mm \widetilde V_n ~=~ \tilde x_{n-1}\, x_1 \,, }
$$
where $x_1$ is the degree 1 ``Casimir'' of $U(1)$.

Set $V_{2j}=v_{2j}$ with $v_{2j}$ being some constant and eliminate
$x_2, \dots, x_{2n-4}$ from the deformed relations $V_{2j}=v_{2j}$
$(1\leq j \leq n-1)$. We are then left with
$$
\tilde x_{n-1}^2+(-1)^n\, x_1^{2(n-1)}
+\sum_{j=1}^{n-1} \hat v_{2j}\, x_1^{2(n-1-j)} ~=~ 0\,,
\hskip10mm   \hat v_{2j} ~=~ (-1)^{n+j}v_{2j} \,,
$$
and $\widetilde V_n -\tilde v_n=0$. Introduce the superpotential so that
$$
\eqalign{
{\del W \over \del x_1}
~=~ & \half\, \big( \tilde x_{n-1}^2+(-1)^n\, x_1^{2(n-1)}
+\sum_{j=1}^{n-1} \hat v_{2j}\, x_1^{2(n-1-j)} \big) \,,  \cr
{\del W \over \del \tilde x_{n-1}}
~=~ & \tilde x_{n-1}\, x_1 -\tilde v_n  \,, }
$$
then the desired $D_{2n}$ potential is obtained as
\eqn\Dsing{W ~=~ \half\, y^2\, x+{(-1)^n \over 2\, (2n-1)}\, x^{2n-1}
+\sum_{j=1}^{n-1} {\hat v_{2j}\over 2\, (2n-2j-1)}\, x^{2n-2j-1}
-\tilde v_n\, y  \,,}
where we have set $x_1=x$ and $\tilde x_{n-1}=y$.

Note that our procedure yields only an $n$-dimensional subset
of the $2n$ possible relevant deformations of the conformal
$D_{2n}$ superpotential:  The foregoing canonical deformation
is {\it odd} under $x \to -x, y\to -y$.   Any other deformations
would destroy the $W(D_n)$ symmetry of the ground states.

We now turn to the $SO(2n)/SU(n)\times U(1)$ coset model, whose
central charge is $c={3n(n-1)\over 2(2n-1)}$.  The $SU(n)$ denominator
factor fixes a (complex) pair of $SO(2n)$ spinors, and so the
ground states of this model are labeled by a set of spinor weights
of $SO(2n)$.  The
\LG variables are, {\it a priori}, the Casimir invariants of  $U(n)$.
Let $b_1, b_2,\dots,b_n$ be the parameters of the $SU(n)$ CSA
with $\sum_{i=1}^n b_i=0$. A set of $SU(n)$ Casimirs can then be
defined by
$$
z_k ~=~ \sum_{1 \leq i_1 < i_2 <\cdots < i_k \leq n}
b_{i_1}\, b_{i_2} \cdots b_{i_k} \,, \hskip10mm k=2,3,\dots,n \,.
$$
In view of the decomposition ${\bf 2n}={\bf n}_2+{\bf {\bar n}}_{-2}$ under
$SO(2n)/SU(n)\times U(1)$, the $SO(2n)$ eigenvalues $a_i$ are expressed
as $a_i=\pm (b_i-2z_1)$, $i=1,2,\dots,n$, where $z_1$ denotes the degree 1
Casimir of $U(1)$. Then one can rewrite the $SO(2n)$ Casimirs \SOcasi\ in
terms
of $z_j$, $j=1,2,\dots,n$.  Inspecting the Casimir decomposition we see that
the variables, $z_j$, of even degree and the variable $z_n$ can be
immediately
eliminated using the vanishing relations.  The $SO(2n)/SU(n)\times U(1)$
model
can thus  be characterized by $[{n \over 2}]$ \LG variables
\foot{Here $[{n \over 2}]$ stands for the integral part of ${n \over 2}$.}
$z_{2j-1}$ of degree $2j-1$ with $j=1,2,\dots,[{n \over 2}]$ and the
superpotential is of degree $2n-1$.  Evaluating the central charge we obtain
the correct value
$$
c ~=~ 3 \sum_{j=1}^{[n/2]} \Big(1-{2\, (2j-1)\over 2n-1} \Big)
~=~ {3n(n-1)\over 2\, (2n-1)} \,.
$$

For $n=2,3,4$ the coset models are identified with the $A_2$, $A_4$ and
$D_8$ minimal models, respectively. It is easily checked that the present
method produces the corresponding \LG superpotentials with
deformations\foot{In particular, the deformed superpotential for the
$SO(8)/SU(4)\times U(1)$ model is converted into that for the
$SO(8)/SO(6)\times U(1)$ model through the $SO(8)$ triality
transformation.}.

The foregoing procedure is, in general, very hard to integrate directly to
obtain the general \LG superpotential, and so we illustrate it in
the first non-trivial example: $n=5$. This model has $c={10\over 3}>3$ and
turns out to be instructive since its shares interesting properties with the
Grassmannian as well as the $E_6$ model. To decompose the Casimirs $V_i$
of $SO(10)$ we take $a_i=-(b_i-2z_1)$.\foot{This choice corresponds to the
${\bf 16}$ of $SO(10)$ as will be seen later. The other choice corresponds
to
the ${\bf \overline{16}}$.}
One finds
\eqn\SOdecom{\eqalign{
V_2(z_j) ~=~ &  20\,z_1^2-2\,z_2 \,,  \cr
V_4(z_j) ~=~ &  160\,z_1^4-16\,z_2\,z_1^2+12\,z_3\,z_1+2\,z_4+z_2^2 \,,  \cr
\widetilde V_5(z_j) ~=~ &  32\,z_1^5+8\,z_2\,z_1^3-4\,z_3\,z_1^2
                        +2\,z_4\,z_1-z_5  \,,   \cr
V_6(z_j) ~=~ &  640\,z_1^6+80\,z_1^3\,z_3-(40\,z_4-12\,z_2^2)\,z_1^2
             -(4\,z_2\,z_3+20\,z_5)\,z_1-2\,z_2\,z_4+z_3^2   \,, \cr
V_8(z_j) ~=~ & 1280\,z_1^8+256\,z_1^6\,z_2+64\,z_1^5\,z_3
               +(48\,z_2^2-160\,z_4)\,z_1^4+(160\,z_5-32\,z_2\,z_3)\,z_1^3
\cr
             & +8\,z_3^2\,z_1^2+(-4\,z_3\,z_4+12\,z_2\,z_5)\,z_1
               -2\,z_3\,z_5+z_4^2  \,, }}
Once again, setting $V_i$ to constant values, $v_i$ ($\tilde v_5$ for
$\widetilde V_5$), we solve the first three equations of \SOdecom\ for
$z_2, z_4$ and $z_5$. We then substitute these values of $z_j$, $j=2,4,5$
into equations for $V_6$ and $V_8$.  With the notation $z_1=x/5, z_3=y$,
the equation for $V_6$ reads
$$
V_6 ~=~ \xcoef(184,625)x^6+\xcoef(144,25)x^3\,y+y^2-\xcoef(38,125)v_2\,x^4
+(-2\,v_4+\xcoef(33,50)v_2^2)\,x^2+(4\,\tilde v_5-\xcoef(4,5)v_2\,y)\,x
+\xcoef(1,2)v_2\,v_4-\xcoef(1,8)v_2^3 \,.
$$
Similarly the equation for $V_8$ becomes
$$
\eqalign{
V_8 ~=~ &  \xcoef(36,625)x^8-\xcoef(24,25)x^5\,y
+4\,x^2\,y^2-\xcoef(12,125)v_2\,x^6
+(\xcoef(6,25)v_4-\xcoef(1,50)v_2^2)\,x^4
+(-\xcoef(56,25)\tilde v_5+\xcoef(4,5)v_2\,y)\,x^3      \cr
&    +(-\xcoef(1,5)v_2\,v_4+\xcoef(1,20)v_2^3)\,x^2
+((\xcoef(1,2)v_2^2-2\,v_4)\,y+\xcoef(6,5)v_2\,\tilde v_5)\,x
+2\,\tilde v_5\,y       \cr
&       -\xcoef(1,8)v_2^2\,v_4
+\xcoef(1,4)v_4^2+\xcoef(1,64) v_2^4  \,. }
$$
In order to obtain the \LG superpotential we set
\eqn\dwdyx{\eqalign{
{\del W \over \del y} ~=~ & \half\, (V_6-v_6) \,,  \cr
{\del W \over \del x} ~=~ & (V_8-v_8)
        +\half\, (\xcoef(16,25)x^2-\xcoef(2,5)v_2)\, (V_6-v_6) \,.}}
The second term of ${\del W \over \del x}$ is introduced so that
the integrability condition is satisfied
$$
{\del \over \del x}\,\Big( {\del W \over \del y}\Big) ~=~
{\del \over \del y}\,\Big( {\del W \over \del x}\Big)  \,.
$$
By integrating \dwdyx\ we obtain the deformed \LG superpotential for
the $SO(10)/SU(5)\times U(1)$ coset model
\eqn\WSOten{\eqalign{
W ~=~ & \xcoef(2372,140625)x^9+\xcoef(92,625)x^6\,y+\xcoef(36,25)x^3\,y^2
+\xcoef(1,6)y^3-\xcoef(788,21875)v_2\,x^7+(\xcoef(63,1250)v_2^2
-\xcoef(2,25)v_4)\,x^5      \cr
& +(-\xcoef(19,125)v_2\,y-\xcoef(6,25)\tilde v_5)\,x^4
+(-\xcoef(8,75)v_6+\xcoef(3,25)v_2\,v_4-\xcoef(61,1500)v_2^3)\,x^3   \cr
& +\big((-v_4+\xcoef(33,100)v_2^2)\,y
+\xcoef(1,5)v_2\,\tilde v_5 \big)\,x^2   \cr
& +(-\xcoef(1,5)v_2\,y^2+2\,\tilde v_5\,y-v_8-\xcoef(9,40)v_2^2\,v_4
+\xcoef(1,4)v_4^2+\xcoef(13,320)v_2^4+\xcoef(1,5)v_2\,v_6)\,x        \cr
& +(\xcoef(1,4)v_2\,v_4-\xcoef(1,16)v_2^3-\xcoef(1,2)v_6)\,y  \,.}}

This procedure does not expose any obviously generalizable
structure, and so it seems rather hard to use it to extract
the superpotential for general $n$.  Fortunately, it is possible
to generalize the algorithm developed for Grassmannians.
The crucial observation is again that the canonical deformation
defines a point in the Cartan subalgebra, while the  ground states
of the  $SO(2n)/(SO(2n-2)\times U(1))$ and $SO(2n)/(SU(n)\times U(1))$
models are respectively characterized in terms of vector and spinor
weights.  We can thus map the \LG potential \Dsing\ onto the
\LG potential for $SO(2n)/(SU(n)\times U(1))$ by finding the
auxiliary equation that relates the vector and spinor weights, and
then cross-eliminating in the proper manner.

We start by eliminating, or integrating out, the variable $y$ in \Dsing\
using ${\del W \over \del y}=0$ to get:
\eqn\Dnelim{ \widetilde W(x) ~=~
{(-1)^n \over 2\, (2n-1)}\, x^{2n-1}
+\sum_{j=1}^{n-1} {\hat v_{2j}\over 2\, (2n-2j-1)}\, x^{2n-2j-1}
-{\tilde v_n^2 \over 2\, x} \,.}
The $2n$ roots of this equation are, by construction, the $2n$
vector weights, $x= \pm a_j$.  The critical points of
the $SO(2n)/(SU(n) \times U(1))$ model are determined
by the $2^{n-1}$ spinorial combinations: $\pm \half a_1 \pm \dots
\pm \half a_n$, with an even (or odd) number of $-$ signs.
The simplest auxiliary equation has roots $x=a_j, j=1,\dots, n$
and so determines the vector weights up to a sign.
One now rewrites this polynomial equation in terms of the $z_j$:
That is, we define the polynomial
$$
\eqalign{
R(x) ~=~ & \prod_{i=1}^n\, (x-a_i) ~=~ \prod_{i=1}^n\, (x+b_i-2\,z_1)  \cr
~=~ & x^n-2\, n\, z_1\, x^{n-1}+\sum_{k=2}^n\, c_k(z_j)\, x^{n-k} \,. }
$$
When calculating the coefficients $c_k(z_j)$ we utilize the deformed
relations
$V_{2i}(z_j)=v_{2i}$ to eliminate $z_{2i}$ ($i=1,2,\dots$) in favor of
$z_{2j-1}$ with $j=1,2,\dots,[{n \over 2}]$. Then we have
\eqn\Rcoefk{c_k(z_j) ~=~ c_k(z_1,z_3,\dots ;\, v_{2j}) \,,
\hskip10mm k=2,3,\dots, n-1 \,, }
whereas the relation $\widetilde V_n(z_j)=\tilde v_n$ yields
\eqn\Rcoefn{c_n(z_j) ~=~ (-1)^n\, \tilde v_n \,. }

For example, for  $n=5$ we find:
$$
\eqalign{
& c_2 ~=~ 50\,z_1^2-\xcoef(1,2)v_2 \,,   \hskip10mm
  c_3 ~=~ -140\,z_1^3+3\,v_2\,z_1+z_3 \,,   \cr
& c_4 ~=~ 150\,z_1^4-5\,v_2\,z_1^2-10\,z_3\,z_1
            -\xcoef(1,8)v_2^2+\xcoef(1,2)v_4 \,, \hskip10mm
  c_5 ~=~ -\tilde v_5 \,.        }
$$

The  \LG superpotential for the $SO(2n)/(SU(n) \times U(1))$ coset model
is then computed by summing $\widetilde W(x)$ over all the roots of
$R(x)=0$:
\eqn\spinpot{W(z_1,z_3,\dots,z_{2[{n\over 2}]-1};\, v_{2j}, \tilde v_n)
~=~ \sum_{i=1}^n \, \widetilde W(a_i) \,,}
where the sums of powers of the $a_j$ that appear on the
right-hand side are evaluated by making use of the auxiliary
equation $R(x)=0$. In particular, the sum of the
last term of $\widetilde W(x)$ becomes
$$
-{\tilde v_n^2 \over 2}\, \sum_{i=1}^n\, {1\over a_i}
~=~ {(-1)^n \over 2}\, \tilde v_n \, c_{n-1}(z_j) \,,
$$
where \Rcoefn\ and \Rcoefk\ have been used.

We have confirmed that this method correctly recovers the deformed
superpotentials obtained by the procedure of linear elimination for
the models with $n=2,3,4$ and $5$.

\subsec{The canonically deformed $E_6$ superpotential}

There is only one SLOHSS model involving each of $E_6$ and $E_7$, and
so there are no short-cuts: we need to use the Casimir decompositions.
Here we present the computation for $E_6$.  The details for $E_7$
are similar, but more complicated, and so they have been included in
an appendix.

The denominator of the $E_6$ model involves $SO(10)$, which fixes
a weight in the {\bf 27} of $E_6$, and there are thus $27$ ground states.
The first step to obtaining the superpotential is to decompose the six
Casimirs, $V_j$, of $E_6$ into the Casimirs, $x_j$, of
$SO(10) \times U(1)$.   This was done explicitly
in \LWpoly, but we reproduce the result here since
there was a typographical error in the printed version.
$$
\eqalign{
V_2({x_j})\ &=\ 12\, {{x_1}^2} + x_2 \,, \cr
V_5({x_j})\ &=\ 48\, {{x_1}^5} - 8\, {{x_1}^3} x_2 +
    x_1 {{x_2}^2} - 2\, x_1 x_4 + 4\, x_5  \,, \cr
V_6({x_j})\ &=\  - 4680\, {{x_1}^6} - 1062\, {{x_1}^4} x_2 -
   \xcoef(177,2) {{x_1}^2} {{x_2}^2} - \xcoef(23,8) {{x_2}^3} -
   15 \, {{x_1}^2} x_4 + \xcoef(5,4) x_2 x_4 \cr
&\ \ \ \  - 60 \,x_1 x_5 - x_6   \,, \cr
V_8({x_j})\ &=\  25830 \,{{x_1}^8} + 7098\, {{x_1}^6} x_2 +
   \xcoef(3027,4) {{x_1}^4} {{x_2}^2} +
   \xcoef(363,8) {{x_1}^2} {{x_2}^3} + \xcoef(171,128) {{x_2}^4} +
   \xcoef(555,2) {{x_1}^4} x_4 \cr
&\ \ \ \  + \xcoef(105,4) {{x_1}^2} x_2 x_4 -
   \xcoef(15,32) {{x_2}^2} x_4 - \xcoef(35,32) {{x_4}^2} +
   1740\, {{x_1}^3} x_5 + 75\, x_1 x_2 x_5 - 6 \,{{x_1}^2}x_6\cr
&\ \ \ \ - \xcoef(1,2) x_2 x_6 + \xcoef(15,8) x_8  \,, \cr}$$
$$\eqalign{
V_9({x_j})\ &=\
28560 \,{{x_1}^9} - 1008\, {{x_1}^7} x_2 + 42\, {{x_1}^5} {{x_2}^2} +
   35\, {{x_1}^3} {{x_2}^3} + \xcoef(105,16) x_1 {{x_2}^4} -
   924\, {{x_1}^5} x_4 \cr
&\ \ \ \  - 70\, {{x_1}^3} x_2 x_4 -
   \xcoef(105,4) x_1 {{x_2}^2} x_4 + \xcoef(35,4) x_1 {{x_4}^2} +
   840 \,{{x_1}^4} x_5 + 420\, {{x_1}^2} x_2 x_5\cr
&\ \ \ \
  + \xcoef(35,2) {{x_2}^2} x_5 - 7\, x_4 x_5 - 112\, {{x_1}^3} x_6 +
   28\, x_1 x_2 x_6 - 21\, x_1 x_8  \,,
\cr}$$
$$\eqalign{
V_{12}({x_j})\ &=\  177660\, {{x_1}^{12}} + 97902\, {{x_1}^{10}} x_2 +
   \xcoef(36063,4) {{x_1}^8} {{x_2}^2} +
   \xcoef(1635,4) {{x_1}^6} {{x_2}^3} +
   \xcoef(7569,64) {{x_1}^4} {{x_2}^4}\cr
&\ \ \ \
  -\xcoef(577,128) {{x_1}^2} {{x_2}^5} - \xcoef(15,1024) {{x_2}^6} +
   \xcoef(15705,2) {{x_1}^8} x_4 + \xcoef(6087,2) {{x_1}^6} x_2x_4-
   \xcoef(7299,16) {{x_1}^4} {{x_2}^2} x_4 \cr
&\ \ \ \
  +\xcoef(1741,32) {{x_1}^2} {{x_2}^3} x_4 +
   \xcoef(1307,1536) {{x_2}^4} x_4 +
   \xcoef(7527,16) {{x_1}^4} {{x_4}^2} -
   \xcoef(795,32) {{x_1}^2} x_2 {{x_4}^2} -
   \xcoef(423,256) {{x_2}^2} {{x_4}^2} \cr
&\ \ \ \
  +\xcoef(85,128) {{x_4}^3} +
   94104\, {{x_1}^7} x_5 + 13050 \,{{x_1}^5} x_2 x_5 +
   \xcoef(1551,2) {{x_1}^3} {{x_2}^2} x_5 -
   \xcoef(59,8) x_1 {{x_2}^3} x_5\cr
&\ \ \ \
   + 219\, {{x_1}^3} x_4 x_5 +
   \xcoef(243,4) x_1 x_2 x_4 x_5 + 6 {{x_1}^2} {{x_5}^2} -
   \xcoef(19,2) x_2 {{x_5}^2} - 948\, {{x_1}^6} x_6 \cr
&\ \ \ \
   + 1041\, {{x_1}^4} x_2 x_6 -
   \xcoef(313,4) {{x_1}^2} {{x_2}^2} x_6 -
   \xcoef(61,48) {{x_2}^3} x_6 - \xcoef(25,2) {{x_1}^2} x_4 x_6 +
   \xcoef(25,24) x_2 x_4 x_6 \cr
&\ \ \ \
  - 50\, x_1 x_5 x_6 +
   \xcoef(1,3) {{x_6}^2} - \xcoef(4257,4) {{x_1}^4} x_8 +
   \xcoef(561,8) {{x_1}^2} x_2 x_8 + \xcoef(97,64) {{x_2}^2}x_8-
   \xcoef(45,32) x_4 x_8  \,.
\cr}
$$
We then solve the equations $V_i = v_i$, where $v_i$ are constants,
for $i=2,5,6,8$.  These are the equations that are linear in $x_i$,
and so we use them to directly eliminate $x_2, x_5, x_6$ and $x_8$.
This leaves the equations: $V_9 = v_9$ and $V_{12} = v_{12}$.
These are integrable, and former is
proportional to ${\del W \over \del x_4}$, and the latter must be
proportional to  ${\del W \over \del x_1} + p_3(x_j, v_j)
{\del W \over  \del x_4}$, where $p_3(x_j, v_j)$ is a homogeneous
polynomial of degree $3$.

For reasons that will become apparent in sections 3.4 and 4.2,
we replace the $v_j$ with some other constants, $w_\alpha$ according to:
\eqn\vwrels{\eqalign{v_2 ~=~ & -2\, w_1 \,, \qquad
v_5 ~=~ -4\, w_2 \,, \qquad v_6 ~=~  w_3 + 20\, w_1^3\,, \cr
v_8 ~=~ & 5 \, w_4 +  4\, w_1 w_3 + 15\, w_1^4\,, \qquad
v_9 ~=~-28\,( w_5+  2\, w_1^2 w_2 )\,, \cr
v_{12} ~=~ & 5\, w_6 +  \coeff{26}{3}\, w_1^2
w_4 + \coeff{1}{3}\, w_3^2 + 4 \, w_1^3  w_3 + 19\, w_1 w_2^2 +
 w_1^6 \,.}}
Finally, setting $x_1 = {1 \over 4} x$ and
$x_4 = {60 \over 13} z + {117 \over 16}x^4  + 5 w_1 x^2  +
2 w_1^2$ one arrives at the superpotential:
\eqn\superWE{\eqalign{ W~=~ & x^{13}   -\coeff{25}{169} \,x\,z^3 +
\coeff{5}{26} \,z^2\,w_2  \cr & +  z\,\Big( x^9 + x^7\,w_1 +
\coeff{1}{3}\, x^5\,{w_1}^2 - x^4\,w_2 - \coeff{1}{3}\,x^2\,w_1\,w_2 +
\coeff{1}{12}\, x^3\,w_3 - \coeff{1}{6}\, x\,w_4 +
\coeff{1}{3} \, w_5 \Big)  \cr & + \coeff{247}{165} \,x^{11}\,w_1 +
\coeff{13}{15} \,x^9\,{w_1}^2 -  \coeff{39}{20} \,x^8\,w_2 +
\coeff{169}{945}\,x^7\,{w_1}^3  + \coeff{13}{105} \,x^7\,w_3 -
\coeff{26}{15} \,x^6\,w_1\,w_2 \cr & +
\coeff{13}{225}\,x^5\,w_1\,w_3 - \coeff{13}{50} \,x^5\,w_4 -
\coeff{91}{180}\,x^4\,{w_1}^2\,w_2 +  \coeff{13}{30}\,x^4\,w_5  +
\coeff{13}{15} \,x^3\,{w_2}^2 -  \coeff{13}{90} \,x^3\,w_1\,w_4 \cr & -
\coeff{13}{120} \,x^2\,w_2\,w_3   + \coeff{13}{90}\,x^2\,w_1\,w_5 -
\coeff{13}{270} \,x\,w_6 - \coeff{13}{360} \,{w_1}^4\,w_2  +
\coeff{13}{90} \,{w_1}^2\,w_5   \,.}}
This potential then has the property that:
\eqn\partW{ {\del W \over \del z} ~=~ \coeff{1}{84}\, (V_9 - v_9)\,;
\qquad {\del W \over \del x} ~=~ \coeff{13}{1350}\,(V_{12} - v_{12})
~+~ \coeff{13}{210}\, (x^3 - \coeff{1}{6}\, w_1 x)\, (V_9 - v_9)\,.}

\subsec{Single variable potentials}

One can always partially eliminate variables from the
\LG potential by using some of the equations, ${\del W \over
\del x_j} =0$.  Indeed we have already done this for all the
quadratic variables in $W$:  this is what is meant by using
the linear vanishing relations.  One can go further and eliminate all
variables except the one of lowest degree, the Casimir, $x$,
of degree $1$. The result, $\cW(x;v_j)$, is defined to be
the single variable potential.  To be more explicit, let
$W(x,y,z,\dots;\, v_i)$ be the deformed  superpotential,
where $x,y,z,\dots$ are the \LG variables. One
can solve the equations of motion
\eqn\elimyzetc{ {\del W\over \del y} ~=~ {\del W\over \del z} ~=~
\dots ~=~ 0\,,}
and expressing $y,z,\dots$ in terms of $x$ and $v_i$, and substitute
the result back into $W$ to obtain $\cW(x;v_j)$.  This one variable
potential has the property that all the ground states are determined from
the solutions of:
$$
{d \,\cW \over dx} ~=~ {d \, \over dx}\,
W(x,\, y_{cl}(x),\, z_{cl}(x),\dots ;\, v_i) ~=~ 0\,.
$$

In general one will not be able to solve \elimyzetc\
explicitly, and so the one variable potential is generically implicit.
On the other hand, for several important examples, the equations
can be explicitly solved and one obtains a polynomial
or irrational potential.

There is also another more direct way to get
the single variable potential, and this approach is more directly
related to the applications considered in \refs{\EMNW,\TESKYa}.  In this
approach one takes the characteristic polynomial, $P(x;v_j)$, of a general
CSA matrix in the representation of $G$ that corresponds to the
ground state of the SLOHSS model of interest.  One then shifts the Casimir
of highest degree according to $v_r \to v_r + \tau$, and solves
$P(x;v_1,\dots, v_{v-1}, v_r + \tau) = 0$ for $\tau(x; v_j)$.
The roots of the equation $\tau(x; v_j)=0$ are precisely the
ground states of the SLOHSS model, and so the single variable potential
in terms of the $U(1)$ Casimir, $x$, is given by:
\eqn\singvarpot{\cW(x; v_j) ~=~ \int~ \tau(x; v_j) ~dx \,.}
We thus refer to $\tau(x; v_j)$ as the pre-potential.
This function is discussed further in the Appendix.

Given an irrational or implicit potential in a single variable, one can
reconstruct  the multi-variable, polynomial potentials in a relatively
straightforward  manner:  One replaces every algebraically
independent irrational or implicit form by a new chiral primary field.
Algebraic vanishing relations are then introduced so that their solution
yields the relationships between the new variables and the original
irrational or implicit forms.
One then integrates these vanishing relations (where possible) to obtain
a \LG potential.  For all the SLOHSS potentials that we have studied in
detail, we have found that the irrational, single variable potentials
are equivalent to  the full algebraic, multi-variable potentials.

There are several reasons why the single variable potentials are useful.
First, we will find that the results of the period integrals in the
$4$-fold will most directly reduce to the single variable
potential.  Knowing the single variable potentials is thus valuable data
in the reconstruction process.   More generally, one finds such single
variable potentials also naturally arise in the construction of
Seiberg-Witten
Riemann surfaces from period integrals on Calabi-Yau $3$-folds
\refs{\EMNW,\SWStrings,\WLNWEsix}.

There is a further application of the single variable potentials
in the coupling of topological matter to topological gravity
\refs{\TopGr,\TESKYa}.  In this context,  the physical operators
and their correlators are most directly expressed in terms of a
single variable potential and residue formulae.

We will therefore illustrate both approaches to computing
the single variable superpotentials.  We will use the first method on
the $E_6$ superpotential, and arrive at results closely related to those
of \refs{\WLNWEsix,\TESKYa}.  We will then use the second approach
to rederive the result from section 3.2 for $SO(2n)/U(n)$.

For $E_6$, one can eliminate the variable $z$ from \superWE\ by solving
${\del W \over \del z} =0$.  One finds  that:
$z = {13 \over 30 x}(w_2 \pm \sqrt{p_2})$ where,
\eqn\ptwo{\eqalign{p_2 ~\equiv&~ 12\,x^{10} + 12\,w_1\,x^8 +
4w_1^2\,x^6 -  12\,w_2\,x^5 + w_3\,x^4 - 4\,w_1 w_2\,x^3  -
2\,w_4\,x^2 \cr &\, + 4\,w_5\,x  + w_2^2 \ .}}
Substituting this into the superpotential one obtains:
\eqn\Wonevar{\cW ~=~ \coeff{13}{270}\,\Big( q_0 ~\pm~ {1 \over 2 x^2}
\,(\sqrt{p_2}\,)^3 \Big) \,,}
where
\eqn\qzero{\eqalign{q_0 ~\equiv&~ \coeff{270}{13} \,x^{13} +
\coeff{342}{11} \,w_1 \,x^{11} +  18\,{w_1}^2 \,x^9 -
\coeff{63}{2}\,w_2  \, x^8  + \coeff{2}{7}\,\big( 13\,{w_1}^3 +
9\,w_3 \big) \,x^7 -   27 \,w_1\,w_2 \,x^6  \cr &  +
\coeff{3}{5} \,\big( 2\,w_1\,w_3 - 9\,w_4 \big) \,x^5 -
\coeff{3}{2} \,\big( 5\,{w_1}^2\,w_2 - 6\,w_5 \big) \,x^4 +
3\,\big( 3\,{w_2}^2 - w_1\,w_4 \big) \,x^3 \cr & -
\coeff{3}{2}\,\big( w_2\,w_3 - 2\,w_1\,w_5 \big) \,x^2
-  \big( 3\,w_1\,{w_2}^2  +   w_6 \big) \,x \cr & -
\coeff{3}{4} \,\big( {w_1}^4\,w_2 + 2\,w_2\,w_4 -
4\,{w_1}^2\,w_5 \big) +  3\,{w_2\,w_5  \over x} +
\coeff{1}{2}\,{{w_2}^3 \over x^2} \,.}}
This is precisely the one-variable $E_6$ potentials discussed in
\refs{\WLNWEsix,\TESKYa, \ItYa}. In particular, one has:
\eqn\derivsW{{d q_0 \over dx} ~=~ {1 \over x^3}\, q_1 -w_6 \,,
\qquad  {d  \over dx}\bigg[
{1 \over 2 x^2} \,(\sqrt{p_2}\,)^3 \bigg]
~=~ {1 \over x^3}\, p_1 \, \sqrt{p_2}\,,}
where
\eqn\pqdefs{\eqalign{
p_1 ~\equiv&~  78\,x^{10} + 60\,w_1 \,x^8 + 14\,w_1^2\,x^6  -
33\,w_2\,x^5 + 2\,w_3 \,x^4 - 5\,w_1 w_2\,x^3 - w_4\,x^2 \cr
&\, - w_5\,x - w_2^2\,, \cr
q_1 ~\equiv&~  270\,x^{15} + 342 \,w_1\,x^{13} +
162\,w_1^2\,x^{11} - 252\,w_2\,x^{10} + (26\,w_1^3 + 18\,w_3)\,x^9 \cr
&\,- 162\,w_1 w_2\,x^8 + (6\,w_1 w_3 - 27\,w_4)\,x^7
 - (30\,w_1^2 w_2  - 36\,w_5)\,x^6 \cr
&\, + (27\,w_2^2 - 9\,w_1 w_4)\,x^5 -  (3\,w_2 w_3 - 6\,w_1 w_5) \,x^4
- 3\,w_1 w_2^2\,x^3\cr
&\, - 3\,w_2 w_5\,x  - w_2^3 \,.}}
The pre-potential:
\eqn\SWpot{\tau ~\equiv~ {d \cW \over dx} ~=~ {1 \over x^3}\,\big(
q_1 ~\pm~  \, p_1 \, \sqrt{p_2} \big)~-~ w_6 \,}
is precisely the one that plays a crucial role
in defining the Seiberg-Witten Riemann surface for $E_6$ \WLNWEsix.

We now start from the other end for the $SO(10)/(SU(5) \times U(1))$ model:
The ground states are classified by the ${\bf 16}$ of $SO(10)$, and
the relevant characteristic polynomial is therefore given by:
$$
P_{SO(10)}^{\bf 16}(x)
=\prod^{16}\, \big(x - (\pm \xcoef(1,2)a_1 \pm \xcoef(1,2)a_2 \cdots
        \pm \xcoef(1,2)a_5) \big)
$$
with an even number of ``$-$'' signs.  Expanding this we have
$$
\eqalign{
P_{SO(10)}^{\bf 16}(x) ~=~ &
{1 \over 16}\, (q_0^2-\xcoef(1,64)q_1^2\,q_2-2\, v_8\,q_0+v_8^2)    \cr
~=~ & x^{16}-2\,v_2\,x^{14}+(\xcoef(7,4)v_2^2-v_4)\,x^{12}-12\, v_5\,x^{11}
      + \cdots  \,,  }
%+ (-\xcoef(7,8)v_2^3+\xcoef(3,2)v_2\,v_4-2\,v_6)\, x^{10}+\cdots \,,    }
$$
where we have put the Casimirs to $V_{2i}=v_{2i}$ and
$\widetilde V_5= \tilde v_5$. Here $q_0, q_1$ and $q_2$ are polynomials
$$
\eqalign{
q_0 ~=~ &  68\,x^8-20\,v_2\,x^6-8\,(-\xcoef(7,4)v_4+\xcoef(5,16)v_2^2)\,x^4
           -24\,\tilde v_5\,x^3-8\,(-\xcoef(3,32)v_2^3+\xcoef(3,8)v_2\,v_4
           -\xcoef(1,2)\,v_6)\,x^2      \cr
        &  +2\,\tilde v_5\,v_2\,x+\xcoef(1,64)v_2^4
        +\xcoef(1,4)v_4^2-\xcoef(1,8)v_2^2\,v_4  \,, \cr
q_1 ~=~ &  48\,x^5-8\,v_2\,x^3-v_2^2\,x+4\,v_4\,x-4\,\tilde v_5  \,,  \cr
q_2 ~=~ &  2\,(64\,x^6-16\,v_2\,x^4+(-4\,v_2^2+16\,v_4\,x^2 )\, x^2
             -32\,\tilde v_5\,x+v_2^3-4\,v_2 \,v_4+8\,v_6)  \,.   }
$$
Note that
\eqn\qrel{{dq_2 \over dx}-16\, q_1 ~=~ 0  \,. }
Then we obtain the pre-potential, $\tau$, by solving
$P_{SO(10)}^{\bf 16}(x; v_8+\tau)=0$:
$$
\tau_{\bf 16}(x) ~=~ -v_8+q_0 \pm {1\over 8}\, q_1\, \sqrt{q_2} \,.
$$
This expression can then be integrated with the use of \qrel\ to give:
$$
\cW(x) \equiv \int ~ \tau_{\bf 16}(x) ~dx   ~=~
\int \, (-v_8+q_0)dx \pm {1\over 192} (\sqrt{q_2})^3 \,.
$$

To get to a multi-variable, rational potential one simply has to replace
$\sqrt{q_2}$ by a new \LG variable.  It is convenient to
mix this with other degree $3$ terms in $x$ and $v_j$ to arrive
at the definition:
\eqn\newvary{ y  ~=~  \pm \sqrt{q_2} ~-~ \coeff{72}{25}\,x^3 ~+~
\coeff{2}{5}\, v_2\, x \,.}
One now replaces all occurences of $\sqrt{q_2}$ in $\cW(x)$ using \newvary,
and the result is precisely $W(x,y)$ of \WSOten.  Moreover,
the equation \newvary\ is exactly the solution of
${\del W\over \del y}=0$.

\newsec{Landau-Ginzburg Models and the IIA theory on $4$-folds}

\subsec{Review}

The construction of \GVW\ starts by identifying the vacuum states of
the IIA theory that correspond to the Landau-Ginzburg vacua.
These IIA vacua are obtained by first choosing a Calabi-Yau
$4$-fold, $Y$, and then choosing a set of fluxes for the
$4$-form field strength, $G$, on $Y$.  There is a quantization
condition on the fluxes which requires that the  quantity:
\eqn\fluxquant{N ~=~ {\chi \over 24} ~+~ \half~
\int~{G \wedge G \over (2 \pi)^2} \,,}
be an integer.  This means that the characteristic class
$\xi \equiv [{G \over 2 \pi}]$ satisfies a shifted Dirac
quantization condition in that the difference between any
two fluxes $G$ and $G^\prime$ must satisfy
$\xi - \xi^\prime \in H^4(Y;\ZZ)$.  The integer $N$ is then
the number of ``world-filling'' fundamental strings in
the vacuum state of the theory, and these fundamental strings
are to be located at some points, $P_i$, chosen in $Y$.
For the vacuum to have a mass gap one must thus have $N=0$.

Domain walls, or solitons are represented by kinks
that interpolate between spatial regions in which the $G$-fluxes
are different.  In particular, for any element $S$ of
the integral homology $H_4(Y,\ZZ)$, we may choose a $D4$-brane
that wraps this cycle and appears as a
``particle'' in the  $\IR^{1,1}$ ``world''.  Such a brane is
a source of $G$-flux, and indeed across this domain wall
one has $\xi - \xi^\prime =[S]$ where  $[S]\in H^4(Y;\ZZ)$
is the Poincar\'e dual of $S$.  Thus the model is specified
by the family of $G$-fluxes obtained that are mapped into
one another via such solitonic $D4$-branes.

Having chosen a $G$-flux, the complex and
K\"ahler structures on
$Y$ are required to satisfy constraints \refs{\Beckers,\GVW}
in order to preserve the requisite supersymmetry with a zero
cosmological constant.  Specifically, if $G_{p,q}$
are the $(p,q)$ parts of $G$ one requires that $G_{0,4} =
G_{1,3} =0$, and  if $K$ is the complexified K\"ahler form,
one requires that $G \wedge K = 0$.  These imply that
$G$ must be a self-dual $(2,2)$-form.  In \GVW\ a superpotential
was proposed for the complex structure and K\"ahler moduli.  We
will only consider the former here, and it is given by:
\eqn\Wconj{W(T_i) ~=~ {1 \over 2 \pi}~\int_Y ~\Omega
\wedge G \,.}
The variation of Hodge structure means that $W$ and $dW$
vanish in the vacuum states, and the result is the
Landau-Ginzburg theory at a conformal point.
More generally, one can seek vacua of massive Landau-Ginzburg
theories, and then one only impose the condition that
$dW=0$ for a vacuum state.

A Landau-Ginzburg soliton is a BPS state whose mass is
equal to its topological charge, and the latter is given
by the change in the value of the superpotential along the
soliton.  Since the solitons are represented by $D4$-branes
wrapping integral $4$-cycles, one thus concludes that:
\eqn\deltaW{\Delta W ~=~ {1 \over 2 \pi} \int_Y~ \Omega
\wedge (G - G^\prime) ~=~ \int_{[S]}~ \Omega \,,}
where $[S]$ is (the  class of) the $4$-cycle dual to
the class $\big[{(G - G^\prime) \over 2\pi}\big]$.

For non-compact Calabi-Yau manifolds, the quantization condition,
\fluxquant, is replaced by a boundary condition
at infinity.  That is, there is a new conserved quantity,
the ``flux at infinity:''
\eqn\fluxinfty{\Phi ~=~ N ~+~ \half~
\int~{G \wedge G \over (2 \pi)^2} \ ,}
and the value of $\Phi$ must be given in order to specify the
model.  Going to non-compact Calabi-Yau manifolds
also affects the dynamics in other non-trivial ways.
In particular, some of the complex structure moduli will
give rise to scalar fields whose kinetic terms are non-normalizable.
These moduli will thus have to have zero kinetic energy, and
thus their dynamics is frozen for the non-compact manifold.  These
complex structure moduli thus become true moduli, or
coupling constants of the \LG model.

In this paper we will follow \GVW\ and consider only non-compact
Calabi-Yau manifolds that are fibrations of
$ALE$ singularities over some complex $2$-dimensional base.
Specifically, we consider non-compact $4$-folds
defined by the equation $P(z_1,\dots,z_5) = 0$ where
$P(z_1,\dots,z_5) = H(z_1,z_2) -  z_3^2
- z_4^2 - z_5^2$ and:
\eqn\Stypes{\eqalign{H(z_1,z_2)~=~ & z_1^{n+1} ~+~ z_2^2  +
\dots  \qquad A_n \,, \cr
H(z_1,z_2)~=~ & z_1^{n-1} ~+~ z_1\, z_2^2  +\dots
\qquad D_n \,, \cr
H(z_1,z_2)~=~ & z_1^3 ~+~ z_2^4  +\dots
\qquad E_6 \,,\cr
H(z_1,z_2)~=~ & z_1^3 ~+~ z_1\, z_2^3  +\dots
\qquad E_7  \,, \cr
H(z_1,z_2)~=~ & z_1^3 ~+~ z_2^5  +\dots
\qquad E_8 \,.}}
In these expressions $+\dots$ indicates the addition of
all possible relevant deformations of the singularity.
The number of monomials involved in these deformations
is the rank of the $ADE$ group, and the natural degrees of
the coefficients of the monomials are the degrees of
the corresponding Casimir invariants.
In \GVW\ it was shown that for these singularities the condition
$G \wedge K=0$ is always satisfied, and {\it all} the
moduli of the singularity are coupling constants, and indeed
have no dynamics.

The homology, $H_4(Y,\ZZ)$, of a surface determined
by \Stypes\ is naturally
identified with the root lattice, $\Gamma$, of the corresponding group,
and the intersection form is the Cartan matrix. The Poincar\'e duals of
these cycles are the elements of the compact cohomology,
$H_{cpct}^4(Y;\ZZ)$, while the full set of non-trivial
$G$-fluxes, $\xi$, are classified by $H^4(Y;\ZZ)$, which
can be identified with the weight lattice, $\Gamma^*$, of the group.
The fluxes are thus characterized by weights of the group,
monodromies of the singularity will permute these by the
action of the Weyl group, and solitons can add or subtract
root vectors.

A model is thus specified by a weight, $\xi$, in $\Gamma^*$.  Conservation
of $\Phi$ means that adding or subtracting a root either acts as a
Weyl reflection, preserving the  magnitude of the flux, or it shortens
the weight, trading some flux for a string in the ground state.
As was argued in \GVW, a state with $N \ne 0$ has massless excitations
coming from motions of the string.  Thus a non-trivial model with a mass
gap must have $N = 0$ in all states, and the weight, $\xi$, must therefore
be  miniscule.  (Non-miniscule weights can be shortened by adding or
subtracting
roots.)  In terms of the ``flux at infinity,''  $\Phi$, the models with a
mass gap
are those with minimum value of $\Phi$ in each of the classes
$\Gamma^*/\Gamma$.

Thus the models with mass gaps are classified by miniscule representations
of the underlying $ADE$ group, and as we saw earlier, the choice of
a miniscule weight determines the denominator group of the coset model, and
leads us to the hermitian symmetric space models \HSSlist.

We now wish to examine how the \LG potentials of the SLOHSS models
emerge from the period integrals of the surfaces defined by \Stypes.
{}From the foregoing discussion there is an obvious problem that will
arise: all the dynamics is frozen.  The singularity, and its period
integrals encode only topological data about the \LG theory,  and do not
contain dynamical fields.  It turns out that we will still be able
to extract the \LG potentials from this topological data, and thus
implicitly find some \LG fields.  While we cannot give
definitive geometric characterizations of these \LG fields, we
find that they emerge in the calculation in a very interesting
and natural manner.     We will also find that the evaluation of the
period integrals has implicit ambiguities
whose resolution corresponds to selecting the miniscule weight, or
flux at infinity, and thus determines the \LG variables and chiral ring.
We will remark further upon this in sections 4.4, and we will begin
by direct evaluation of periods.

\subsec{Integrating over $4$-cycles}

The integral of the $4$-form, $\Omega$, over a cycle, $S$,
of the non-compact surfaces defined by \Stypes\ can be written:
\eqn\intOmS{\int_{[S]} \ \Omega ~=~ \int \ {dz_1\, dz_2\, dz_3\,
dz_4 \over  \sqrt{ H(z_1,z_2)- z_3^2 - z_4^2 } }   \,,}
where the $4$-cycle is defined as follows:  $z_3$ runs around
branch-cut of the square-root; $z_4$ runs between
$z_4 =\pm  \sqrt{H(z_1,z_2)}$, {\it i.e.} points at which
the branch cut shrinks to a point; the variables $z_1, z_2$
are then integrated over a $2$-cycle, $S_2$, of the singularity
$ H(z_1,z_2) - z_5^2  = 0$.  The first two integrals are
elementary, and reduce to the following (up to an overall
normalization):
\eqn\interint{\int_{S_2}\ dz_1\, dz_2 \, \sqrt{H(z_1,z_2)}\,.}
For the $A_n$ singularity, one has $H(z_1,z_2) = P(z_1;a_j) +
z_2^2$,  where $P(z_1; a_j)$ is a polynomial of degree $n+1$.
As was discussed in \GVW, the integral \interint\ then reduces
to:
\eqn\Atypeint{\int \ dz_1\, P(z_1;a_j)\,,}
evaluated between the zeroes of $P(z_1;a_j)$.  That is, we have
recovered the superpotential, W, of \Wbasic, with the
integral ({\it i.e.} topological charge of the
soliton)  expressed in terms of the values of $W$ at its
critical points.

It is almost as elementary to perform the integrals for the
$D_n$ singularity.  In this instance one has:
$H(z_1,z_2) = P(z_1;a_j) + z_1 z_2^2 +  a_0 z_2$, where
$P(z_1;a_j)  = z_1^{n-1} + \sum_{j=0}^{n-2} a_{n-1-j} z_1^j$.
One makes the change of variable $z_1 = x^2$, and one performs
the elementary integral over $z_2$ to obtain:
\eqn\Dtypeint{\cW ~=~ \int\,dx \,\Big(\, P(x^2;a_j)~-~
{a_0^2 \over 4\, x^2} \, \Big) ~=~  \int \,  P(x^2;a_j) \,dx ~+~
{a_0^2 \over 4\, x }   \,,}
which is to be evaluated between the zeroes of the integrand.

This is the single variable potential \Dnelim\  for the
${SO(2n)\over SO(2n-2)\times U(1)}$ coset model. As we remarked earlier,
it is obtained from \Dsing\ by using ${\del W \over \del y} =0$ to
eliminate  $y$. The superpotential, \Dsing,  can be recovered
by removing the singular term by introducing the variable
$y = {a_0 \over 2 x}$.  Also observe that \Dtypeint\ is an
odd function and contains only half of the general versal
deformations of the $D_n$ singularity.

The calculation for the $E_6$ singularity is essentially the
same as that performed in \WLNWEsix, and so we will only sketch
the details here.

One starts with the singularity:
\eqn\cubic{\eqalign{H_{E_6}(z_i) ~=~ & z_1^3 ~+~ z_2^4
 ~+~ \coeff{1}{2}\, w_1~z_1\, z_2^2 ~-~
\coeff{1}{4}\,w_2\, z_1 \,z_2 ~+~ \coeff{1}{96}
(w_3 - w_1^3 )\, z_2^2   \cr &
~+~\coeff{1}{96} \big(w_4 + \coeff{1}{4} \, w_1\, w_3 -
\coeff{1}{8}\,w_1^4 \big)\,  z_1 ~-~ \coeff{1}{48}
\big(w_5 - \coeff{1}{4}\,w_1^2\, w_2 \big)\, z_2    \cr &
~+~ \coeff{1}{3456}\, \big(\, \coeff{1}{16} \, w_1^6  -
\coeff{3}{16} \, w_1^3\,w_3  +
\coeff{3}{32}\, w_3^2 - \coeff{3}{4} \,w_1^2\,w_4 + w_6
\,\big)\, ,} }
where we have made a convenient choice for the
deformation parameters, $w_j$.
One reparameterizes the singularity by setting
\eqn\reparam{z_1 ~=~ x\, y + \alpha(x) \, ;
\qquad z_2 ~=~ y + \beta(x) \, ,}
and the   result is a quartic in $y$.  One now sets
\eqn\abchoices{\eqalign{\beta(x)  ~=~ & -\coeff{1}{4}\,
\big(\, x^3 + \coeff{1}{2}\,w_1 \,x \,\big)\,, \cr
 \alpha(x) ~=~ & \coeff{1}{48}\,(2\,w_1\,x^2 +
\,w_1^2) ~+~ \coeff{1}{24\, x}\,
 \big(\,w_2 ~\pm~ \sqrt{p_2} \, \big) \,,}}
where $p_2$ is given by \ptwo.  The functions $\beta$
and $\alpha$ are chosen so that the $y^3$ and
$y^1$ terms in the quartic vanish respectively.
The integral \interint\ now takes the form:
$$
\int\,dx\, dy\,(y + \alpha'(x) - x\, \beta'(x))\,
\sqrt{ y^4 + A(x) \,y^2 + B(x) } \,,
$$
for some functions, $A(x)$ and $B(x)$.  By taking a
contour at large $y$ this integral reduces to:
\eqn\intres{\eqalign{\int\,dx\,\big(\,\coeff{1}{4}\,
(A(x))^2 ~-~  B(x) \,\big) & ~=~  \coeff{1}{3456 }\,
\int\,dx \,\big[\,\coeff{1}{x^3}\,
(q_1 \pm p_1\,\sqrt{p_2}\,) - w_6  \,\big] \cr & ~=~
\coeff{1}{3456 }\,  \Big( q_0 ~\pm~ {1 \over 2 x^2}
\,(\sqrt{p_2}\,)^3 \Big)  \,,}}
and thus we regenerate the one variable superpotential
\Wonevar\ with all the correct moduli.  As we remarked earlier,
one can reconstruct \superWE\ by replacing the singular
irrational part using:
$$
{13 \over 30 x}\,(w_2 ~\pm~ \sqrt{p_2}) \equiv z \,.
$$

\subsec{Other superpotentials}

Our integration procedure appears to have led us directly
to a single  superpotential for the $A_n$ and
$D_n$ models.  The reason for this is that we must have
implicitly chosen a flux, or boundary condition at
infinity.  Indeed, the sleight-of-hand occurred when
we passed from the definite integral to the indefinite
integral with a single end-point to the integration.

We know from the algorithms of section 3 that we can get
the multi-variable potentials for any of the Grassmannian
models and for the $SO(2n)/U(n)$ series by summing the
potentials \Wbasic\ and \Dsing\ over carefully selected
{\it sets} of their critical points.  Exactly the same choice
emerges in the period integrals.  For example, in the
Grassmannian we could take the end-points of the integration
to be a subset of $m$ solutions of ${dW \over dx} =0 $, with
this subset defined by \xzreln.  The result would be an indefinite
integral that depends upon $z_1, \dots, z_m$, and reproduces
the multi-variable potential for the Grassmannian.

We therefore see that the selection of the $G$-flux
in $H^4(Y,\ZZ)$, or the choice of the flux at infinity, amounts
to selecting a set of  critical points of the single
variable potentential. In the next subsection we will demonstrate
how this works more explicitly by showing that the ``boundaries''
of the integration procedure, and in particular these
critical points, correspond to special holomorphic divisors
of the surfaces defined by \Stypes.

It is also interesting to note that if we now consider
the period integral defined by two critical points
of the Grassmannian potential, $W_{m,n}(z_1,\dots,z_m)$,
then it will decompose into a difference  of sums of
critical values of the superpotential $W_{1,m+n-1}(x)$.
Generically this will represent the sum of topological charges
of a multi-soliton (and thus non-BPS) state.  Thus not all
pairs of ground states represent boundary conditions for
BPS solitons.  This fact was first discovered
by considering the integrable \LG models, where it was seen that
one could not make a factorizable $S$-matrix involving only
solitons for certain classes of integrable model.  Some boundary
conditions could only give rise to multi-soliton states
\refs{\IntLG,\LWpoly}.  Indeed, based on the conserved quantities of the
$E_6$ model it was shown in \LWpoly\ that for the $E_6$ integrable
model, the correct soliton spectrum was exactly given by
the prescription that two ground states labeled by weights
$\lambda_1$ and $\lambda_2$ would give rise to a single, fundamental BPS
soliton {\it if and only if} $\lambda_1- \lambda_2$ was exactly
a root.  It is now very satisfying to have a simpler, and far more general
string theoretic explanation of this result:  A wrapped of $D4$-branes on
the ADE  singularity is a fundamental BPS soliton if and only if it
wraps a cycle that is represented by a root of the Lie algebra.

\subsec{Intersection forms and holomorphic $4$-cycles}

We have just seen how the \LG potentials emerge from some
kind of  ``semi-periods'' of the singularity:  That is, we get the
potentials by making the indefinite integral of
the single variable potential, and then parameterizing families
of its critical points.  This suggests that one should try to
characterize these families of critical points and
the ``semi-periods'' more geometrically.  As we will see, this is
indeed possible since they represent non-compact homology cycles of
the singularity.  Ideally we would like to integrate $\Omega$ over these
non-compact cycles, but such an integral will diverge.  This can
be fixed by compactifying the singularity.
Our approach is based upon the techniques used in
section 17 of \NSEWb, and we will therefore only summarize the key steps
here.

The first step is to
introduce a new coordinate and make the equation
of the singularity homogeneous (see  \NSEWb\ for details).
The resulting compact manifold will {\it not} be a Calabi-Yau manifold
and so $\Omega$ will have some kind of problem at infinity:
If one wants to preserve holomorphicity, then
it will be singular.   Since we wish to consider
period integrals, we wish to keep $\Omega$ regular, but
the cost is the loss of holomorphy in the patch
at infinity.   This will not affect the computation of
topological charges since they only involve  integrals
of $\Omega$ over compact cycles of the original singularity.
This means that such a compactification will only affect
the \LG superpotential by some additive constant.

The basic idea now is to try to write   $\Omega = d\,\lambda$
for some $3$-form, $\lambda$.  This is not possible precisely
because $\Omega$ is a non-trivial element of cohomology.
However, if one excises all the non-trivial homology from
the compactified singularity then it is possible to write
$\Omega = d\,\lambda$.  Let $C_a$ be a basis of the non-trivial
$4$-cycles.  One can then write
$$
\Omega = d\,\lambda ~+~ \, \sum_a\, m_a\, [C_a]
$$
where $[C_a]$ denotes the Poincar\'e dual of the cycle
$C_a$, and is a $4$-form with delta-function support
on $C_a$.  The parameters, $m_a$ are determined from
the integral of $\Omega$ over $C_a$:  Specifically, if
$M_{ab}  = {\cal I}(C_a,C_b)$ is the intersection form of the
basis of cycles, then:
$$
m_a  ~=~  \sum_b \, M^{-1}_{ab}\, \int_{C_b} \, \Omega \,.
$$
Given any other cycle, $S$, one then has:
$$
\int_S \, \Omega ~=~ \sum_a \, m_a\  {\cal I} (S,C_a) \,,
$$
where ${\cal I} (S,C_a)$  denotes the intersection numbers
of the cycle $S$ with $C_a$.

The issue now is to find a good basis, $C_a$; and there is
a particularly nice way to do this using holomorphic lines
in the ADE singularity in two complex dimensions, or using
holomorphic planes in four complex dimensions.    The important
point is that $\Omega = {dz_1 \, dz_2 \, dz_3 \, dz_4 \over
z_5}$  is {\it odd} under any reflection $z_j \to - z_j$,
and so any holomorphically defined dual homology cycle
must similarly be odd.

In practice, given the singularity type:
\eqn\singform{z_5^2 ~=~ H(z_1, z_2) ~-~ z_3^2 ~-~z_4^2 \,,}
one starts by looking for complex ``$2$-planes'' of the form
\eqn\planes{ z_1 ~=~ a\, \zeta + b \,, \qquad  z_2 ~=~ c\,
\zeta + d \,, \qquad z_3 ~=~ \pm \, i\, z_4 \,,}
where $\zeta$ is a complex variable, and $a, b, c$ and $d$
are constants.  For generic $a, b, c, d$ \singform\ and
\planes\ define an irreducible rational surface that is even
under $z_5 \to - z_5$.  At special values of $a, b, c, d$ one finds
that $ H(a\, \zeta + b,  c\,  \zeta + d) $ is a perfect square
and the rational surface degenerates to two intersecting planes:
The odd $4$-cycles that we seek are the differences of
such pairs of $2$-planes\foot{Technically
they will not strictly be $2$-planes since $z_5$ is generically
going to be polynomial in the other variables.}.  More
generally, one can seek holomorphic $2$-surfaces of higher
degree (see, for example, \Entori).  Such holomorphic surfaces
form Weyl orbits whose order depends upon the degree of
the surface.  Here we are focussing on ``planes,'' or surfaces
of lowest degree so as to recover the miniscule orbits.  One could
also go to surfaces of higher degree, and these are presumably
relevant to the physics of more general fluxes at infinity.

Going back to the earlier parts of this section, one sees that
the planes described above played a crucial role in
the explicit evaluation of the integrals.  The cycles
were defined by families of contour integrals and the planes
defined where these contours collapse to points, that
is, the planes defined the limits of the integration.
In simple terms, a $4$-cycle is essentially an $S^4$ and
its intersection points with $2$-planes define the extremities,
or ``north'' and  ``south'' poles of the $S^4$.  Thus the
difference of two lines, $C_a - C_b$ defines
the compact cycle over which we integrate, and the
fact that integration is the difference  in values of the
superpotential at the two extremities simply reflects
the fact that the integral is given by
$m_a - m_b$ and that $m_d$ is the value of the superpotential
at the line $C_d$.

The Grassmannian and $SO(2n)/U(n)$ models are obtained by considering
families of critical points of the simplest  potentials, {\it i.e.}
families of critical points for either $U(n)/(U(n-1) \times U(1))$ or
$SO(2n)/(SO(2n) \times U(1))$, and then summing these simple potentials
over  such families.  This means that the more general superpotentials
must correspond to taking periods of {\it sums of} holomorphic $2$-surfaces
in the singularity.  The correspondence with the flux at infinity
is now much more transparent: The Poincar\'e duals of these surfaces
can be taken to be $\delta$-functions of their volume forms.
Since the surfaces are holomorphic, their volume forms are necessarily
$(2,2)$-forms.   We thus see a very explicit correspondence
between the ``non-compact''  $(2,2)$-fluxes of the singularity,
and the ground states of the \LG superpotential.

Having compactified the singularity, we might hope to have
unfrozen the dynamics and recover some information about the
dynamical \LG fields.  In compactifying we have made some choices
about how to regularize $\Omega$ at infinity.  We expect these
choices to be related to the choice of the ``irrelevant'' $D$-terms,
and the unrenormalized dynamical \LG variables should emerge
in terms of properties that can be expressed entirely in terms
of the non-compact singularity.  Indeed, by examining the holomorphic
$2$-surfaces  more explicitly for  the examples above,
we find a much more direct role for the Landau-Ginzburg
variables.

For the $A_n$ singularity the zeroes of $P(z_1;a_j)$
are precisely the points where the singularity contains
the planes:
$$
 z_5 ~=~ \pm\,i\, z_2\,, \qquad z_3 ~=~ \pm \, i\, z_4 \,.
$$
For the $D_n$ singularity the solutions of $P(x^2;a_j)=
{a_0^2 \over 4 x^2}$ define precisely the points
where the singularity contains the planes:
$$
z_5 ~=~ \pm\, x \, \big(\,z_2 ~+~  {a_0  \over 2 x^2} \,\big)
\,, \qquad z_3 ~=~ \pm \, i\, z_4 \,.
$$
Indeed, more generally, the $D_n$ superpotential is given by
$W(x,y) \equiv \int P(x^2;a_j) dx  -  x\, y^2 + a_0\, y$,
and reduces to \Dtypeint\ upon eliminating $y$ via: ${\del W
\over \del y} =0$.  The critical points of the {\it full}
superpotential $W(x,y)$ precisely define the planes:
$$
z_5 ~=~ \pm\, (\,x \, z_2 ~+~  y \, )
\,, \qquad z_3 ~=~ \pm \, i\, z_4 \,,
$$
lying in the $D$-type singularity.

The story is similar for the $E_6$ and $E_7$ singularities.
The details for the $E_7$ singularity may be found in
Appendix A.  For the $E_6$, the surfaces are quadratic in $z_5$,
but linear in the other $z_j$.  Once again take $z_3 ~=~ \pm \,
i\, z_4$ and introduce $\zeta \in \IC$ with
$$
\eqalign{
z_1 ~=~ &   x \, \zeta  ~+~  \coeff{5}{52}\, z ~+~
\coeff{1}{48} (2\, w_1\, x^2 + w_1^2) \,, \cr
z_2 ~=~ &  \zeta   ~-~  \coeff{1}{8}(2\, x^3 + w_1\,x )
\,,}
$$
The parameters of these surfaces are $x$ and $z$.  When the
superpotential $W(x,z)$ of \superWE\  has a critical point then
the foregoing defines a ``plane'' in the $E_6$  singularity \cubic\ with:
$$
z_5 ~=~\pm\, \big(\, \zeta^2 + \coeff{3}{16}\,x^6 + \coeff{1}{8}
\,w_1\, x^4 + \coeff{5}{192}\,w_1^2  \,x^2- \coeff{1}{8} \,w_2\, x +
  \coeff{1}{192}\,w_3 ~+~ \coeff{15}{104}(x^2 + \coeff{1}{6}\,w_1) \,
z \,\big)  \,.
$$
Indeed, the $27$ critical points of $W(x,z)$ corresponding to
the weights of the fundamental of $E_6$ define the celebrated
27 lines in a cubic hypersurface in $\IP^3$.

We therefore see that the Landau-Ginzburg variables naturally
emerge in the parameterization of representatives of the non-compact
homology cycles.   In retrospect this is rather natural:  we know that
the  \LG variables must parameterize a  family of fluxes, and
have non-normalizable kinetic terms in the non-compact
singularity.  Dual to such fluxes are non-compact
homology $4$-cycles, and so the \LG dynamics can be converted
into a description of these $4$-cycles: they intersect the
compact cycles of the singularity in a manner determined by the inner
products of the corresponding weight and root vectors, and
the dynamics is such that these supersymmetric
ground states correspond   holomorphic  surfaces.
It is therefore tempting to think of the ground-state flux
as being created by a non-compact $D4$-brane that threads the
singularity along one of these non-compact cycles.  The solitons
then intersect these non-compact branes and then combine with
them to yield another non-compact $D4$-brane that threads
the compact part of the singularity with different set of
intersection numbers.

\newsec{Conclusions}

The $\cN=2$ superconformal models that arise from the non-compact
ADE  singularities are determined by a flux at infinity.
For the minimal, or miniscule, fluxes the versal deformations
of the singularity lead to a theory with a mass gap.
It was argued in \GVW\ that these particular models should
be perturbations of the SLOHSS models.  Here we observed
that the perturbations of the singularity respect the
underlying Weyl symmetry, and thus the corresponding
perturbations of the \LG superpotential must do the same.
We then used this to either explicitly construct, or
provide a precise computational algorithm for the
construction of the \LG superpotentials
of the SLOHSS models with  the most general perturbations
consistent with the Weyl symmetry.  We then showed that
exactly the same superpotentials and algorithms
emerged from the topological data and period integrals
of corresponding $ADE$ singularities.  This provides
some extremely detailed checks on the results
of \GVW.   There are several interesting by-products
of our work which may find application in the study
of \LG models and in topological matter coupled to
topological gravity.

There are also quite a number of interesting open questions.
Our work suggests an interpretation of the \LG variables
in terms of moduli of non-compact cycles that thread the
singularity.  It would be interesting to verify these ideas
more explicitly by considering nearly singular, compact Calabi-Yau
manifolds in more detail.

Here we have considered only the theories with a mass gap.
It would be interesting to find out exactly what perturbed
conformal theories emerge for non-miniscule fluxes.  It is tempting
to conjecture that the flux at infinity, $\Phi$, encodes
the level of the numerator current algebra of the coset model.
Since  versal deformation of the singularity is not supposed
to yield a mass gap, the corresponding perturbed \LG models
will have to remain multi-critical.  This is certainly
possible, but there is a potentially interesting conundrum here:
We know from the results in \DVV\ that simple perturbative
attempts at adding marginally relevant operators can create
non-trivial higher order corrections, and indeed result in
a theory with a mass gap.  For example, the $\cN=2$
minimal models perturbed by the least relevant chiral
primary results in a theory with a Chebyshev superpotential.
Indeed, the such a perturbation would be a natural candidate
for the $A_1$ singularity with higher flux.  It would thus
be interesting to see if the resulting theory really does
have massless modes, or if, somehow the theory does indeed
develop a Chebyshev superpotential and hence have a mass gap.
It is also quite possible that the conformal field theories
associated with the singularities with non-miniscule fluxes
are not the various coset models at higher levels and
with different denominators.

One way to approach this problem that might be particularly
interesting would be to find a method of computing the
elliptic genus of the conformal field theory using only
the topological data of the singularity and its flux at
infinity.  For \LG theories this is equivalent to finding
the fundamental \LG fields and determining their $U(1)$
charges.  For more general theories, the elliptic genus
gives a lot of valuable information about the partition
function.

Finally, there are some interesting properties of
\LG solitons that may have interesting consequences
for $D4$-branes more generally.  Most notable, but probably
least relevant is the fact that some very special
perturbations lead to quantum integrable models, with
factorizable $S$-matrices.  This sort of property is
not at all robust, and so it likely to be only a property
of the field theory in the near singular limit.  Integrability
will very likely be spoiled in the full string theory.

More interesting is the fact that solitons have
fractional fermion number \refs{\FracFerm,\PFKI}.  In particular,
for a \LG soliton running between two vacua, the fermion
number is given by \PFKI:
\eqn\fermionnumb{f ~=~ -{1\over 2\pi}\, \Delta\, Im\big(
 \log\big( \det (\del_i\del_j W) \big)\big) \,,}
where $\Delta$ indicates the change in the value of
the quantity between the two critical points.

Note that this quantity is, in principle, a very
complicated function of the moduli and can presumably
assume arbitrary (even irrational) values.  However, for singularities
with discrete geometric symmetries (like the Coxeter
resolution), this fermion number will take simple
fractional values.  One might naturally wonder
whether this fractional fermion number might also be an
artifact of the field theory that emerges near the
singularity, and that this fermion number
would disappear in the full string theory.  There are no global
conserved currents in string theory, and so one might
expect that fermion number would not persist in string theory.
However, the formula   \fermionnumb\ only defines the fermion
number mod $1$, and indeed fermion  number mod $2$ {\it is}
a well defined concept in string theory.  The fact that \fermionnumb\
follows from an index calculation gives some further hope
that fractional fermion number might persist in string theory.
Finally, it seems particularly likely that fractional fermion
number will persist if the fractional fermion number is fixed
by discrete geometric symmetries of the singularity.

If some version of fractional fermion number persists in
string theory then it would be very interesting to find the
geometric meaning of  \fermionnumb.  As regards physical consequences:
there are the obvious phases that would arise in scattering,
but perhaps more interesting would be the consequences
for quantization conditions under compactification, and thus for
partition functions.

%%%%%%%%%%%%%%%%%%%%%%%%%%%%%%%%%%%%%%%%%

%%%%%%%%%%%%%%%%%%%%%%%%%%%%%%%%%%%%%%%%%%%
\bigskip
\leftline{\bf Acknowledgements}

N.P.W. would like to thank S. Gukov and E. Witten for helpful
conversations.  The research of T.E. and S.K.Y. was supported in part
by Grant-in-Aid for  Scientific Research  on Priority Area 707
``Supersymmetry and  Unified Theory of Elementary
Particles'', Japan Ministry of Education, Science and Culture.
T.E. and S.K.Y. are grateful to Japan Society for Promotion of Science
for the exchange program between Tokyo/Tsukuba and USC.
The research of N.P.W. was supported in part
by funds provided by the DOE under grant number DE-FG03-84ER-40168
and by funds under the U.S. Japan Cooperative Science Program:
``String Theory and Field Theory Dualities and Integrable Model;''
NSF Award \#9724831.

\appendix{A}{The $E_7$ Singularity and the Deformed Coset Model}

In this appendix we begin by deriving the single-variable version of the
\LG potential for the $E_7$ singularity closely following \WLNWEsix.
We then obtain the multi-variable \LG potential for the deformed
$E_7/(E_6 \times U(1))$ coset model using the method developed in the
text.   Finally we discuss the integral over 4-cycles
in a non-compact $4$-fold with a singularity of type $E_7$.

\subsec{ The single-variable $E_7$ superpotential}

The $E_7$ singularity with versal deformations is described by
\eqn\defeq{W_{E_7}(z_1,z_2,z_3) ~=~ 0 \,,}
where
\eqn\Esevsing{\eqalign{
W_{E_7}(z_1,z_2,z_3) ~=~ z_1^3+z_1\, z_2^3+z_3^2  &  -w_2 \, z_1^2\, z_2
-w_6\,  z_1^2-w_8 \, z_1\, z_2-w_{10} \, z_2^2   \cr
&  -w_{12}\, z_1-w_{14}\, z_2-w_{18} \,,}}
where $w_q$ are the deformation parameters. To obtain the single-variable
potential we determine the lines for the generic $E_7$ singularity.
The process is very similar to that of the $E_6$ singularity.
We first make a change of variables from $(z_1, z_2)$ to $(x,y)$, where:
\eqn\ytoxy{z_1 ~=~ x^2\, y+\alpha(x), \hskip10mm z_2 ~=~ y+\beta(x) \,,}
where $\alpha(x)$ and $\beta(x)$ will be fixed momentarily.
With these substitutions, $W_{E_7}$  becomes a quartic in $y$.
We then set $\alpha(x)=-x^6+w_2\, x^4-3\, \beta(x)\, x^2$
so as to eliminate the $y^3$ term in this quartic.  Eliminating the term
linear in $y$ gives rise to a cubic equation for $\beta(x)$. It is
convenient to introduce a function $Y(x)$, where
$\beta(x)={Y(x) \over 4x}+x^4$, and then
the cubic for $Y(x)$ reads
\eqn\Ycubic{Y^3+3\, q(x)\, Y-2\, r(x)~=~0 \,,}
where $q(x)$ and $r(x)$ are polynomials in $x$ given by:
\eqn\qrpol{\eqalign{
 q ~=~ & -28\, x^{10}-\xcoef(44,3) v_2\, x^8-\xcoef(8,3) v_2^2\, x^6
-\xcoef(1,3) v_6\, x^4+\xcoef(2,9) v_8\, x^2-\xcoef(1,3) v_{10} \,, \cr
 r ~=~ & 148\, x^{15}+116\, v_2\, x^{13}+36\, v_2^2\, x^{11}
+(4\, v_2^3+\xcoef(8,3) v_6)\, x^9+(\xcoef(2,3) v_2v_6-2\, v_8)\, x^7  \cr
    & +(2\, v_{10}-\xcoef(2,3) v_2v_8)\, x^5
+(\xcoef(1,81) v_6^2-\xcoef(2,27) v_{12})\, x^3
      +(\xcoef(109,8613) v_2v_6^2-\xcoef(2,3) v_{14})\, x \,.  }}
In writing the foregoing we have reparameterized the deformations of
the $E_7$ singularity as follows:
\eqn\wtov{\eqalign{
& w_2~=~-v_2 \,, \quad w_6~=~\xcoef(1,12) v_6 \,, \quad
w_8~=~-\xcoef(1,6) v_8 \,, \quad w_{10}~=~-\xcoef(1,4) v_{10} \,, \cr
& w_{12}~=~\xcoef(1,108) (2\, v_{12}-\xcoef(1,3) v_6^2) \,, \quad
w_{14}~=~\xcoef(1,12) (2\, v_{14}-\xcoef(109,2871) v_2\, v_6^2) \,, \quad
w_{18}~=~-\xcoef(1,36) v_{18}  \,.  }}
Equation \defeq\  now reduces to:
\eqn\Wafter{z_3^2 ~=~ -(x^2y^4+A(x)\, y^2+B(x)) \,.}

The ``lines'' are determined by requiring that the right-hand
side of \Wafter\ be a perfect square, which means
 that the discriminant, $\Delta$, vanishes:
\eqn\discri{\eqalign{
0 ~=~ \Delta ~=~ & A^2-4\, x^2 B   \cr
~=~ & \xcoef(1,16) (p_1\, Y^2+p_2\, Y+p_3-\xcoef(16,9)v_{18}\, x^2) \,.}}
The polynomials $p_1$, $p_2$ and $p_3$ are
\eqn\ppoly{\eqalign{
p_1 ~=~ & 1596\, x^{10}+88\,v_2^2\, x^6+7\, v_6\, x^4+660\, v_2\, x^8
          -2\,v_8\, x^2 -v_{10}\,,   \cr
p_2 ~=~ & 16872\, x^{15}+11368\, v_2\, x^{13}+2952\, v_2^2\, x^{11}
          +(176\, v_6+264\, v_2^3)\, x^9  \cr
        & +(-100\, v_8+\xcoef(100,3)v_2\, v_6)\, x^7
          +(-\xcoef(68,3)v_2\, v_8+68\, v_{10})\, x^5  \cr
        & +(\xcoef(2,9)v_6^2 -\xcoef(4,3)v_{12})\, x^3
          +(\xcoef(218,8613)v_2\, v_6^2-\xcoef(4,3)v_{14})\, x   \,,  \cr
p_3 ~=~ & 44560\, x^{20}+41568\, v_2x^{18}+16080\, v_2^2\, x^{16}
          +(2880\, v_2^3+\xcoef(2216,3)v_6)\, x^{14}  \cr
        & +(312\, v_2\, v_6+192\, v_2^4-\xcoef(1552,3)v_8)\, x^{12}
          +(32\, v_2^2\, v_6-40\, v_2\, v_{10}-\xcoef(64,3)v_{12}
            +\xcoef(11,3)v_6^2)\, x^8   \cr
        & +(-\xcoef(416,3)v_{14}-16\, v_2^2\, v_{10}-\xcoef(4,9)v_6\, v_8
            -\xcoef(32,9)v_2\, v_{12}+\xcoef(27776,8613)v_2\, v_6^2)\, x^6
\cr
        & +(\xcoef(3488,8613)v_2^2\, v_6^2+\xcoef(4,9)v_8^2
           -\xcoef(64,3)v_2\, v_{14}-\xcoef(2,3)v_6\, v_{10})\, x^4
          +\xcoef(4,3)v_8\, v_{10}\, x^2+v_{10}^2 \,.
}}
The polynomials $q$, $r$, $p_1$, $p_2$ and $p_3$ play an important role in
our calculations. In particular they obey remarkable identities
\eqn\polyident{
x\, {dq\over dx}~=~\half \, q-{1\over 6}\, p_1\,, \hskip10mm
x\, {dr\over dx}~=~{3\over 4}\, r+{1\over 8}\, p_2 \,.  }
The condition $\Delta =0$ yields a single-variable version of the
pre-potential, $\tau_{E_7}(x; v_i)$, for $E_7$,
\eqn\Esevsingl{
\tau_{E_7}(x; v_i)~=~ -v_{18}+{9\over 16\, x^2}\, (p_1\, Y^2+p_2\, Y+p_3)
\,, }
where $Y$, a root of the cubic \Ycubic, is given by
$$
Y ~=~ \{ s_+ + s_- \,, \quad  \omega\, s_+ + \omega^2\, s_- \,, \quad
         \omega^2\, s_+ + \omega \, s_- \}
$$
with $\omega = e^{2\pi i/3}$ and $s_\pm = (r\pm \sqrt{q^3+r^2})^{1/3}$.
Using the relation between the roots and the coefficients of the cubic
equation it is easy to verify from \Esevsingl\ that
\eqn\taucubic{
(\tau_{E_7}+v_{18})^3+ A_2(x)\, (\tau_{E_7}+v_{18})^2
+A_1(x)\, (\tau_{E_7}+v_{18}) +A_0(x) ~=~0 \,, }
where
$$
\eqalign{
A_2 ~=~ & \xcoef(9,16\, x^2)\, (6\, q\, p_1-3\, p_2)  \,,  \cr
A_1 ~=~ & (\xcoef(9,16\, x^2))^2 \, (9\, q^2\, p_1^2-6\, r\, p_1\, p_2
           -12\, q\, p_1\, p_3+3\, q\, p_2^2+3\, p_3^2)  \,, \cr
A_0 ~=~ & -(\xcoef(9,16\, x^2))^3 \, (4\, r^2\, p_1^3 +6\, q\, r\, p_1^2\,
p_2
            +9\, q^2\, p_1^2\, p_3-6\, r\, p_1\, p_2\, p_3
            -6\, q\, p_1\, p_3^2   \cr
        &   \hskip18mm  + 2\, r\, p_2^3+3\, q\, p_2^2\, p_3+p_3^3)  \,.
}
$$

The single-variable pre-potential, $\tau_{E_7}(x;v_i)$,
can be integrated with respect to $x$ to yield the
single variable potential, ${\cal W}_{E_7}(x;v_i)$. This
integral has two parts:
$$
{\cal W}_{E_7}(x;v_i) ~=~ \int dx\, \tau_{E_7}(x;v_i) ~=~
\cI_1 +\cI_2 \,,
$$
where
$$
\cI_1 ~=~ \int dx\, \Big(-v_{18}+{9\over 16\, x^2}\,p_3 \Big) \,, \hskip10mm
\cI_2 ~=~ {9\over 16} \int dx\, {1\over x^2} (p_1\, Y^2+p_2\, Y) \,.
$$
The integral $\cI_1$ is easily performed. We only note that since $p_3$
has no linear term in $x$ there appears no logarithm of $x$. To evaluate
$\cI_2$, on the other hand, we first use \polyident\ to derive
$$
\cI_2 ~=~ {9\over 16} \int dx\, \left(
{d\over dx} \Big(-{6\, q\over x}\, Y^2+{8\, r\over x}\, Y \Big)
+(3\, q\, Y-2\, r)\,
\Big( {4\over x}\, {dY \over dx}-{Y\over x^2}\Big) \right) \,.
$$
By virtue of \Ycubic\ the second term is reduced to
$$
(3\, q\, Y-2\, r)\,
\Big( {4\over x}\, {dY \over dx}-{Y\over x^2}\Big)
~=~ -{d \over dx} \left( {Y^4 \over x}\right) \,.
$$
Hence we find
\eqn\tauint{
\int dx\, \tau_{E_7}(x;v_i) ~=~ -{27 \over 16\, x}(q\, Y^2-2\, r\, Y)
+\int dx\, \Big(-v_{18}+{9\over 16\, x^2}\,p_3 \Big) \,. }

Before moving on to the multi-variable potential, we wish to
describe how this single-variable potential is related to the
characteristic polynomial of the ${\bf 56}$ of $E_7$.
Generally the characteristic
polynomial of a representation $\cR$ of the Lie algebra $G$ is defined by
$$
P_G^{\cR}(x; V_i) ~=~ \det \, (x-{\vec a}\cdot \vec H)  \,
$$
which is of degree $\dim\,\cR$ in $x$. Here ${\vec a}$ is an $r$-dimensional
vector in the Cartan subspace spanned by $\vec H$ and the $V_i$
$(i=1,\dots, r={\rm rank}\, G)$ are the Casimirs built out of $\vec a$
among which $V_r$ is the top Casimir whose degree equals the Coxeter
number $h$ of $G$. Let $\vec\lambda$ be the weight of $\cR$, then we have
$$
P_G^{\cR}(x; V_i) ~=~
\prod_{\vec\lambda \in \cR}\, (x-\vec a \cdot \vec\lambda) \,.
$$
For the ${\bf 56}$ of $E_7$ we obtain
\eqn\Esevfund{\eqalign{
P_{E_7}^{\bf 56}(x; v_i)~=~ &
x^{56}+ 12\, v_2\, x^{54}+66\, v_2^2\, x^{52}
+(2\, v_6+220\, v_2^3)\, x^{50}   \cr
&  +(10\, v_8+495\, v_2^4+20\, v_2\, v_6)\, x^{48}  \cr
&  +(-126\, v_{10}+792\, v_2^5+90\, v_2^2\, v_6+84\, v_2\, v_8)\, x^{46} \cr
&  +(10\, v_{12}+924\, v_2^6+240\, v_2^3\, v_6
     +\xcoef(934,3)v_2^2\, v_8+86\, v_2\, v_{10})\, x^{44}+ \cdots \,,
}}
where the Casimirs have been set to $V_i=v_i$.
Note that this has the degree 3 in the top Casimir $v_{18}$, which is
analogous to \taucubic. In fact, one can show explicitly that
$$
P_{E_7}^{\bf 56}(x; v_i)~=~ -{x^2\over 3^6}\,
(v_{18}^3+ A_2(x)\, v_{18}^2 +A_1(x)\, v_{18} +A_0(x))  \,.
$$
Hence, \taucubic\ for $\tau_{E_7}$ may be expressed as
$$
P_{E_7}^{\bf 56}(x;\, v_2,\dots,v_{14}, v_{18}+\tau_{E_7})~=~0  \,.
$$
This also makes it clear that each of the lines in the $E_7$ singularity
we have constructed is in correspondence with a weight of
the ${\bf 56}$ of $E_7$.

More generally, for any $ADE$ group, the single-variable \
version of the \LG potential can be obtained from  $\int \tau(x) dx$ ,
where $\tau(x)$ is obtained by solving
$$
P_G^\cR (x;\, v_1,\dots ,v_r+\tau(x)) ~=~0  \,.
$$
Thus the pre-potential, $\tau$, depends upon the representation, $\cR$,
 and takes the form\foot{Although $\tau_\cR (x)$
depends on the representation, it turns out that the description of $ADE$
topological matter theories in terms of $\tau_\cR$ is independent of $\cR$,
but depends only on the singularity type $G$ \TESKYa. This is deeply
related with the universality of the special Prym variety in the theory
of spectral curves of periodic Toda lattice, which plays a fundamental role
in formulating Seiberg-Witten solution of four-dimensional $\cN =2$
Yang-Mills
theory \refs{\EMNW,\ItYa}.}:
$$
\tau_\cR (x; v_i)=-v_r+F_\cR (x;\, v_1,\dots, v_{r-1}),
$$
where $F_{\cR}$ is some irrational, or implicit function of $x$ whose
expansion at $x=\infty$ starts with the $x^h$ term and does not
carry the pole term ${1 \over x}$ as can be seen by degree counting.

\subsec{ The \LG potential for the deformed
$E_7/(E_6\times U(1))$ coset model}

The calculation is also parallel to that for the $E_6/(SO(10)\times U(1))$
model.  We start with the decomposition of the seven Casimirs,
$V_j$, of $E_7$ into the Casimirs $x_j$ of $E_6\times U(1)$.
Under $E_7 \supset E_6\times U(1)$, the ${\bf 56}$ of $E_7$ branches as
$$
{\bf 56} ~=~ {\bf 27}_1+{\bf\overline{27}}_{-1}+{\bf 1}_3+{\bf 1}_{-3} \,.
$$
Accordingly the characteristic polynomial for the ${\bf 56}$ of $E_7$ is
factorized
$$
P_{E_7}^{\bf 56}(x) ~=~ P_{E_6}^{\bf 27}(x+x_1)\cdot
P_{E_6}^{\bf\overline{27}}(x-x_1)\cdot (x+3\, x_1)\cdot (x-3\, x_1) \,,
$$
{}from which we read off the Casimir decomposition as follows:
$$
\eqalign{
V_2(x_j)  ~=~& x_2-3\,x_1^2  \,, \cr
V_6(x_j) ~=~& -60\,x_5\,x_1+x_6-12\,x_2^2\,x_1^2
               -24\,x_2\,x_1^4-72\,x_1^6 \,,  \cr
V_8(x_j) ~=~& 18\,x_2\,x_5\,x_1-24\,x_2\,x_1^6+x_8-12\,x_5\,x_1^3
                -54\,x_1^8+3\,x_6\,x_1^2  \,, \cr
V_{10}(x_j) ~=~&  x_6\,x_1^4-4\,x_2\,x_5\,x_1^3+12\,x_2\,x_1^8
                +4\,x_2^2\,x_1^6+x_5^2+12\,x_1^{10}  \cr
&               -12\,x_5\,x_1^5+4\,x_9\,x_1-2\,x_8\,x_1^2 \,, \cr
V_{12}(x_j) ~=~& -12\,x_2^3\,x_1^6+12\,x_6\,x_1^6+102\,x_2\,x_5\,x_1^5
        +36\,x_2\,x_1^{10}+114\,x_2^2\,x_1^8+x_{12}+\xcoef(1,6) x_6^2   \cr
&       +792\,x_5\,x_1^7-35\,x_2\,x_6\,x_1^4+24\,x_2^4\,x_1^4+270\,x_1^{12}
        -135\,x_8\,x_1^4+132\,x_2^2\,x_5\,x_1^3    \cr
&       +222\,x_5^2\,x_1^2-144\,x_9\,x_1^3-18\,x_2\,x_8\,x_1^2
        -18\,x_2\,x_9\,x_1-4\,x_2^2\,x_6\,x_1^2-2\,x_5\,x_6\,x_1  \,, }
$$
$$
\eqalign{
V_{14}(x_j) ~=~& \xcoef(5402,319)x_2^3\,x_1^8+\xcoef(2616,319)x_6\,x_1^8
        -18\,x_2\,x_5\,x_1^7-\xcoef(10338,319)x_2\,x_1^{12}
        -\xcoef(11358,319)x_2^2\,x_1^{10}   \cr
&       -\xcoef(164616,319)x_5\,x_1^9
        +\xcoef(109,5742)x_6^2\,x_2+x_5\,x_9-\xcoef(436,957)x_2^3\,x_6\,x_1^2
        -\xcoef(66357,319)x_5^2\,x_1^4-x_{12}\,x_1^2  \cr
&       +x_2\,x_6\,x_1^6-\xcoef(76950,319)x_1^{14}
        -\xcoef(2180,957)x_2\,x_5\,x_6\,x_1-2\,x_5\,x_8\,x_1
        +\xcoef(872,319)x_2^4\,x_1^6   \cr
&       -\xcoef(10634,319)x_2^2\,x_5\,x_1^5-36\,x_9\,x_1^5+9\,x_8\,x_1^6
        +\xcoef(17972,319)x_2\,x_5^2\,x_1^2-14\,x_2\,x_9\,x_1^3   \cr
&       +\xcoef(8720,319)x_2^3\,x_5\,x_1^3+x_2\,x_8\,x_1^4
        -\xcoef(109,1914)x_6^2\,x_1^2
        +\xcoef(436,957)x_2^2\,x_6\,x_1^4    \cr
&       +\xcoef(1542,319)x_5\,x_6\,x_1^3+\xcoef(872,319)x_2^5\,x_1^4   \,,
\cr
V_{18}(x_j) ~=~& x_9^2+252\,x_1^{18}+396\,x_2\,x_1^{16}+288\,x_2^2\,x_1^{14}
        +144\,x_5\,x_1^{13}+144\,x_2^3\,x_1^{12}+288\,x_2\,x_5\,x_1^{11}
\cr
&       +24\,x_2\,x_6\,x_1^{10}+216\,x_2^2\,x_5\,x_1^9+21\,x_2^2\,x_6\,x_1^8
        +114\,x_2\,x_8\,x_1^8+72\,x_2^3\,x_5\,x_1^7    \cr
&       -24\,x_5\,x_6\,x_1^7+24\,x_2\,x_9\,x_1^7-12\,x_6\,x_1^{12}
        +36\,x_2^4\,x_1^{10}+168\,x_8\,x_1^{10}-156\,x_9\,x_1^9     \cr
&       +225\,x_5^2\,x_1^8
        +4\,x_6^2\,x_1^6
        +36\,x_{12}\,x_1^6+108\,x_5^3\,x_1^3+x_8^2\,x_1^2
        +12\,x_5\,x_2\,x_8\,x_1^3    \cr
&       -9\,x_5^2\,x_6\,x_1^2-2\,x_{12}\,x_5\,x_1
        +2\,x_8\,x_9\,x_1
        +12\,x_2^2\,x_8\,x_1^6+180\,x_2\,x_5^2\,x_1^6   \cr
&       +84\,x_5\,x_8\,x_1^5
        +12\,x_2^2\,x_9\,x_1^5-4\,x_6\,x_8\,x_1^4+102\,x_5\,x_9\,x_1^4
        +12\,x_2\,x_5\,x_6\,x_1^5     \cr
&       +36\,x_2^2\,x_5^2\,x_1^4
        +10\,x_{12}\,x_2\,x_1^4-4\,x_6\,x_9\,x_1^3
        +12\,x_2\,x_5\,x_9\,x_1^2  \,.  }
$$

Set $V_i=v_i$ with the $v_i$ being arbitrary parameters. Since $V_j$ is
linear
in $x_j$ for $j=2,6,8,12$, we can express as
$x_j = x_j(x_1, x_5, x_9; v_\ell)$ by solving $V_j=v_j$. Substituting these
$x_j$ into $V_k$ with $k=10,14,18$ we are left with three relations
\eqn\Vkvk{V_k(x, y, z; v_\ell)-v_k ~=~ 0,
\hskip10mm k=10,14,18 \,. }
where, for convenience, we have set  $x=3\, x_1$, $y=x_5$, $z=x_9$.
As we found for $E_6$, these relations are integrable. In order to construct
the superpotential, particular linear combinations of the relations \Vkvk\
must be chosen so that they can be integrated to a superpotential.  These
linear combinations are determined using the integrability conditions:
${\del \over \del x} \big( {\del W \over \del y}\big) =
{\del \over \del y} \big( {\del W \over \del x}\big)$, {\it etc.}.
We then find that there is a superpotential, $W(x,y,z)$, with:
$$
\eqalign{
{\del W \over \del z} ~=~
& {3 \over 2}\, (V_{10}-v_{10}) \,, \qquad
{\del W \over \del y} ~=~
  3\, (V_{14}-v_{14})+(\xcoef(43,9)x^4+4\, v_2\, x^2)\, (V_{10}-v_{10}) \,,
\cr
{\del W \over \del x} ~=~
& (V_{18}-v_{18})+(\xcoef(26,27)x^4+\xcoef(4,3)v_2\, x^2)\, (V_{14}-v_{14})
\cr
& +(\xcoef(173,243) x^8+\xcoef(172,9) x^3\,y-v_8+\xcoef(16,9) v_2^2\, x^4
  +\xcoef(508,243) v_2\, x^6+8\, v_2\, x\, y+\xcoef(2,3)v_6\, x^2)\,
(V_{10}-v_{10}) \,,
}
$$
 The explicit form of the superpotential reads:
\eqn\Esevpot{\eqalign{
W ~=~ &
\xcoef(1016644,817887699)x^{19}+\xcoef(33326,177147)x^{14}\,y
+\xcoef(266,6561)x^{10}\,z+\xcoef(16850,2187)x^9\,y^2
+\xcoef(80,27)x^5\,y\,z+\xcoef(124,9)x^4\,y^3      \cr
& +x\,z^2+\xcoef(3,2)y^2\,z
+\xcoef(753964,81310473)v_2\,x^{17}+\xcoef(45392,1594323)v_2^2\,x^{15}
+(\xcoef(18566,2302911)v_6+\xcoef(96064,2302911)v_2^3)\,x^{13}          \cr
& +\xcoef(48826,59049)v_2\,x^{12}\,y
+(\xcoef(6532,216513)v_2\,v_6+\xcoef(1816,72171)v_2^4
-\xcoef(4640,216513)v_8)\,x^{11}
+\xcoef(962,729)v_2^2\,x^{10}\,y        \cr
& +(-\xcoef(173,2187)v_{10}+\xcoef(239,6561)v_2^2\,v_6
-\xcoef(178,2187)v_2\,v_8)\,x^9
+\big((\xcoef(464,729)v_6+\xcoef(418,729)v_2^3)\,y
+\xcoef(110,729)v_2\,z \big)\,x^8               \cr
& +(\xcoef(904,81)v_2\,y^2-\xcoef(508,1701)v_2\,v_{10}
-\xcoef(508,5103)v_2^2\,v_8-\xcoef(4,1701)v_{12}
+\xcoef(23,1701)v_6^2)\,x^7    \cr
& +\big((\xcoef(109,243)v_2\,v_6-\xcoef(145,81)v_8)\,y
+\xcoef(44,243)v_2^2\,z \big)\,x^6      \cr
& +(\xcoef(140,27)v_2^2\,y^2-\xcoef(26,135)v_{14}
+\xcoef(1736,387585)v_2\,v_6^2-\xcoef(2,405)v_2\,v_{12}
-\xcoef(16,45)v_2^2\,v_{10}-\xcoef(29,405)v_8\,v_6)\,x^5        \cr
& +\big((-\xcoef(41,27)v_2\,v_8-\xcoef(43,9)v_{10})\,y
+\xcoef(7,54)v_6\,z \big)\,x^4   \cr
& +(\xcoef(11,9)v_6\,y^2+\xcoef(16,9)v_2\,y\,z-\xcoef(2,9)v_6\,v_{10}
+\xcoef(1,9)v_8^2-\xcoef(4,9)v_2\,v_{14}
+\xcoef(218,25839)v_2^2\,v_6^2)\,x^3            \cr
& +\big(4\,v_2\,y^3+(-\xcoef(1,3)v_{12}-4\,v_2\,v_{10}+\xcoef(1,18)v_6^2)\,y
-\xcoef(1,3)v_8\,z \big)\,x^2+(-v_8\,y^2+v_8\,v_{10}-v_{14})\,x         \cr
& +(-3\,v_{14}+\xcoef(109,1914)v_2\,v_6^2)\,y-\xcoef(3,2)v_{10}\,z   \,.}}
Putting $v_i=0$ and making a change of variables
$x=X_1,\, y=2\, ({2791\over 19})^{1/4}X_5+{416\over 81}X_1^5, \,$
$z=6\, ({2791\over 19})^{1/2}X_9-{593188\over 6561}X_1^9
-{496\over 27} ({2791\over 19})^{1/4} X_1^4 X_5\,$,
we obtain the superpotential at criticality
$$
W ~=~ \xcoef(100476,19) \big( X_1^{19}+X_1\,X_9^2+X_5^2\,X_9
+37\,(\xcoef(19,2791))^{3/4}X_1^{14}\,X_5
-21\,(\xcoef(19,2791))^{1/2}X_1^{10}\,X_9 \big)
$$
which agrees with the result of \LWpoly.

The multi-variable superpotential $W(x,y,z)$ does indeed
reduce to the single-variable potential, $\cW_{E_7}(x)$
of \tauint\ by eliminating $y$ and $z$ from $W$ with the
aid of the equations of motion. First, solving
${\del W \over \del z}=0$ we find:
$$
\eqalign{
z_{cl} ~=~ - {1\over x}\, ( & \xcoef(133,6561)x^{10}+\xcoef(40,27)x^5\,y
+\xcoef(3,4)y^2+\xcoef(55,729)v_2\,x^8+\xcoef(22,243)v_2^2\,x^6
+\xcoef(7,108)v_6\,x^4  \cr
&  +\xcoef(8,9)v_2\,x^3\,y-\xcoef(1,6)v_8\,x^2-\xcoef(3,4)v_{10}) \,. }
$$
Substituting this into ${\del W \over \del y}$ and letting
$y=Y+\xcoef(416,81)x^5+\xcoef(32,27)v_2\, x^3$ lead to
$$
\Big[{\del W \over \del y}\Big]_{z=z_{cl}}
~=~ -{9\over 4\, x}\, (Y^3+3\, q\, Y-2\, r) \,.
$$
Thus ${\del W \over \del y}=0$ is equivalent to
the cubic equation \Ycubic. After some algebra we finally arrive at:
$$
\Big[ {\del W \over \del x} \Big]_{z=z_{cl},y=y_{cl}}
~=~ \tau_{E_7}(x; v_i) \,.
$$
Therefore,
\eqn\tauintW{\int dx\, \tau_{E_7}(x; v_i)~=~
W\big(x,\, y_{cl}(x),\, z_{cl}(x);\, v_i \big) \,. }

\subsec{Integrating over 4-cycles in the $E_7$ singularity}

We consider a non-compact 4-fold defined by \Stypes\ of type $E_7$, where
$H(z_1, z_2)=W_{E_7}(z_1, z_2, z_3=0)$, and evaluate the period integral
\intOmS. As for $E_6$, after the change of variables \ytoxy,
$H(z_1, z_2)$ takes the form of \Wafter, and hence the integral \interint\
becomes
$$
\int dx\,\oint dy\, (2\,x\,y+\alpha '(x)-x^2\,\beta\, '(x))\,
\sqrt{x^2\,y^4+A(x)\,y^2+B(x)} \,.
$$
Performing the integral over $y$ along a contour at large $|y|$ we obtain
$$
\int dx\, \big( \xcoef(1,4\,x) (A(x))^2-B(x) \big)
~=~ -\xcoef(1,36) \int dx\, \tau_{E_7}(x; v_i) \,,
$$
where $\tau_{E_7}(x)$ is given by \Esevsingl.
The remaining integral has already been evaluated. The result is \tauint\
and \tauintW.

In section 4.4 we have seen how holomorphic $2$-surfaces of the form
\planes\ lying in the $A_n$, $D_n$ and $E_6$ singularities are determined
by the critical points of the deformed superpotentials.  Turning to
$E_7$ we may take:
\eqn\Esevpl{\eqalign{
z_1 ~=~ & x^2\,\zeta
+\xcoef(3,133)\big(1787\,x\,y+243\,{y^2\over x^4}+324\,{z\over x^3} \big)
\cr
& +\xcoef(1,133) \big(\xcoef(527,9)v_2\,x^4+88\,v_2^2\,x^2+63\,v_6
+864\,v_2\,{y\over x}-162\,{v_8\over x^2}-729\,{v_{10}\over x^4} \big) \,,
\cr
z_2 ~=~ & \zeta-\xcoef(23,81)x^4+\xcoef(1,4){y\over x}
-\xcoef(8,27)v_2\,x^2 \,, }}
where $\zeta \in \IC$ and $(x,y,z)$ are parameters.
The superpotential $W(x,y,z)$ of \Esevpot\ for the $E_7/(E_6\times U(1))$
model has 56 critical points corresponding to the ${\bf 56}$ of $E_7$.
At a critical point, \Esevpl\ and $ z_3=\pm\, i\,z_4$ describe the
holomorphic $2$-surfaces,  \ie\ $H(z_1,z_2)$
with \Esevpl\ becomes a perfect square, yielding
$$
\eqalign{
z_5 ~=~ \pm \, \big( & x\,\zeta^2-\xcoef(869,26244)x^9-\xcoef(19,27)x^4\,y
+\xcoef(1,4)z-\xcoef(211,2916)v_2\,x^7-\xcoef(1,18)v_2^2\,x^5
-\xcoef(11,432)v_6\,x^3     \cr
& -\xcoef(1,3)v_2\,x^2\,y+\xcoef(1,24)v_8\,x
-\xcoef(1,16){v_{10}\over x} \big) \,.}
$$

%%%%%%%%%%%%%%%%%%%%%%%%%%%%%%%%%%%%%%%%%
% End
%%%%%%%%%%%%%%%%%%%%%%%%%%%%%%%%%%%%%%%%%
\listrefs
\vfill
\eject
\end